%% file: mac_queueing.tex
\documentclass[12pt, draftclsnofoot, onecolumn]{IEEEtran}
\newlength{\figurewidth}
\usepackage[pdftex]{graphicx}
\graphicspath{{./figures/}}
\usepackage[usenames,dvipsnames]{xcolor}

\usepackage{amsmath,amssymb,amsthm,amsfonts}
\usepackage{nicefrac}
\usepackage{cite}
\usepackage{color}
\usepackage{transparent}

\usepackage[caption=false,font=footnotesize]{subfig}

\allowdisplaybreaks

\input{commands}

\IEEEoverridecommandlockouts 
\begin{document}

\title{NOMA in the Uplink: Delay Analysis with Imperfect CSI and Finite-Length Coding}

\author{Sebastian~Schiessl,~\IEEEmembership{Student Member,~IEEE,}
Mikael~Skoglund,~\IEEEmembership{Fellow,~IEEE,}
and~James~Gross,~\IEEEmembership{Senior~Member,~IEEE}
\thanks{The authors are with the School of Electrical Engineering and Computer Science, KTH Royal Institute of Technology, Stockholm, Sweden (e-mail: schiessl@kth.se, skoglund@kth.se, james.gross@ee.kth.se).}
}
\date{\today}
\maketitle

\newtheorem{result}{Result}
\newtheorem{remark}{Remark}
\newtheorem{assumption}{Assumption}
\newtheorem{corollary}{Corollary}

\input{Abstract}
\input{Introduction}
\input{System_Model}

\input{Analysis}

\input{imp_csi}

\input{finite_bl}

\input{Numerics}

\input{Conclusions}

\bibliographystyle{IEEEtran}
\bibliography{../../PHD2}

\end{document}

%% file: commands.tex
\newcommand{\hlchange}[1]{#1}
\newcommand{\hlchangeb}[1]{#1}
\newcommand{\hlremove}[1]{}

\DeclareMathOperator*{\argmin}{arg\,min}
\DeclareMathOperator*{\argmax}{arg\,max}

\newcommand{\defined}{{\;\overset{\Delta}{=}\;}}

\newcommand{\expected}[1]{\mathbb{E}\left[#1\right]}

\newcommand{\Prob}[1]{\mathbb{P}\left\{#1\right\}} 

\newcommand{\Mellin}{\mathcal{M}}

	\newcommand{\Kernelk}[2]{\mathcal{K}_{#1}\left(#2\right)}

\newcommand{\pv}{{p_\mathrm{v}}}
	\newcommand{\pvk}{{p_{\mathrm{v},k}}}
\newcommand{\Abit}{\mathit{A}}
\newcommand{\Dbit}{\mathit{D}}
\newcommand{\Sbit}{\mathit{S}}

\newcommand{\Delay}{{\mathit{W}}}
\newcommand{\Asnr}{\mathcal{A}}

\newcommand{\Ssnr}{\mathcal{S}}

\newcommand{\Xsnr}{\mathcal{X}}

	\newcommand{\ntk}[1]{n_{\mathrm{t},#1}}
\newcommand{\ntotal}{n}
\newcommand{\nd}{n_\mathrm{d}}

\newcommand{\snravg}{\bar{\gamma}}

\newcommand{\snr}{{\gamma}}
\newcommand{\Snr}{{\Gamma}}
\newcommand{\snrmmse}{{\hat{\gamma}}}

\newcommand{\hmmse}{\hat{h}}

\newcommand{\Herr}{\tilde{H}}

\newcommand{\rate}{r}
\newcommand{\rateadapt}{\Phi}
	
	\newcommand{\rateadaptvec}{\boldsymbol{\Phi}}
\newcommand{\opt}{^*}

\newcommand{\VWR}{\eta}

\newcommand{\Npoints}{N^2}

\newcommand{\pvmax}{\delta}
\newcommand{\pvmaxtilde}{\delta_\mathrm{M}}
\newcommand{\pvmaxtildealt}{\tilde{\delta}}
\newcommand{\rateconst}{c}

\newcommand{\snrvec}{{\snr_1,\snr_2}}
\newcommand{\snrveci}{{\snr_{1,i},\snr_{2,i}}}

\newcommand{\sinr}{{\mathsf{SINR}}}
\newcommand{\relaxedprob}{\mathsf{CP}}
\newcommand{\snravgoma}{\bar{\gamma}^\mathrm{o}}

\newcommand{\RIhat}{{\widehat{\rateadapt}_{1}}}

\newcommand{\RIi}{{\rate_{1,i}}}
\newcommand{\RIIi}{{\rate_{2,i}}}

\newcommand{\Rsum}{\rateconst_\Sigma}
\newcommand{\Rsumi}{\rateconst_{\Sigma,i}}
\newcommand{\RImin}{{\rateconst_{1}^\mathsf{min}}}
\newcommand{\RImax}{{\rateconst_{1}^\mathsf{max}}}
\newcommand{\RIImin}{{\rateconst_{2}^\mathsf{min}}}
\newcommand{\RIImax}{{\rateconst_{2}^\mathsf{max}}}
\newcommand{\RImini}{{\rateconst_{1,i}^\mathsf{min}}}
\newcommand{\RImaxi}{{\rateconst_{1,i}^\mathsf{max}}}
\newcommand{\RIImini}{{\rateconst_{2,i}^\mathsf{min}}}
\newcommand{\RIImaxi}{{\rateconst_{2,i}^\mathsf{max}}}

\newcommand{\s}{{\theta}}

\newcommand{\ts}{{t}}

\newcommand{\snrturn}{\gamma_{1,\mathrm{turn}}}
\newcommand{\powfac}{\rho}
\newcommand{\snrfactor}{\beta}
\newcommand{\constc}{\kappa}
\newcommand{\sigman}{\sigma_\mathrm{n}}
\newcommand{\mun}{\mu_\mathrm{n}}
\newcommand{\epsone}{\varepsilon_{1}}
\newcommand{\epstwo}{\varepsilon_{2}}

\newcommand{\epstwodir}{\varepsilon_{2\backslash 1}}
\newcommand{\epstwodirlower}{\varepsilon_{2\backslash 1,\mathrm{a}}}
\newcommand{\epstwodirupper}{\varepsilon_{2\backslash 1,\mathrm{b}}}
\newcommand{\epsonenointerf}{\varepsilon_{1,\mathrm{SIC}}}

\newcommand{\xvec}{\mathbf{x}}

\newcommand{\sigmazk}{\sigma_{Z,k}}

\newcommand{\sigmaicsik}{\sigma_{k}}
\newcommand{\sigmaicsione}{\sigma_{1}}
\newcommand{\sigmaicsitwo}{\sigma_{2}}

\newcommand{\Ptk}{\snravg_{\mathrm{t},k}}

\newcommand{\Pk}{\snravg_k}

\newcommand{\dispersionawgn}{\mathcal{V}_\mathrm{AWGN}}
\newcommand{\dispersioniid}{\mathcal{V}_\mathrm{iid}}
\newcommand{\dispersionmac}{\mathcal{V}_\mathrm{III}}
\newcommand{\fblequiv}{\mathrm{FBL}}
\newcommand{\fblequivone}{{\fblequiv,1}}
\newcommand{\fblequivtwo}{{\fblequiv,2}}
\newcommand{\fblequivonenew}{{\fblequiv,\mathrm{I}}}
\newcommand{\fblequivtwonew}{{\fblequiv,\mathrm{II}}}
\newcommand{\fblequivcomb}{{\fblequiv,\mathrm{III}}}
\newcommand{\Ubone}{U_{1}}
\newcommand{\Ubtwo}{U_{2}}
\newcommand{\sigmafblone}{\sigma_{\fblequiv,1}}
\newcommand{\sigmafbltwo}{\sigma_{\fblequiv,2}}
\newcommand{\sigmaicsifblone}{\sigma_{\mathrm{IC},\mathrm{F},1}}

\newcommand{\Gicsifbltwo}{G_{\mathrm{IC,F,int},2}}
\newcommand{\sigmanewicsifbltwo}{\sigma_{\mathrm{IC,F,int},2}}

%% file: Abstract.tex

\begin{abstract}
We study whether using non-orthogonal multiple access (NOMA) in the uplink of a mobile network can improve the performance over orthogonal multiple access (OMA) when the system requires ultra-reliable low-latency communications (URLLC).
To answer this question, we first consider an ideal system model with perfect channel state information (CSI) at the transmitter and long codewords, where we determine the optimal decoding orders when the decoder uses successive interference cancellation (SIC) and derive closed-form expressions for the optimal rate when joint decoding is used.
While joint decoding performs well even under tight delay constraints, NOMA with SIC decoding often performs worse than OMA.
For low-latency systems, we must also consider the impact of finite-length channel coding, as well as rate adaptation based imperfect CSI. 
We derive closed-form approximations for the corresponding outage or error probabilities and find that those effects create a larger performance penalty for NOMA than for OMA. Thus, NOMA with SIC decoding may often be unsuitable for URLLC.
\end{abstract}

\begin{IEEEkeywords}
Nonorthogonal multiple access (NOMA), stochastic network calculus, effective capacity, quality of service, delay performance, URLLC, imperfect CSI, finite blocklength regime
\end{IEEEkeywords}

%% file: Introduction.tex

\section{Introduction}
\label{sec:introduction}
Non-orthogonal multiple access (NOMA) is considered a viable solution for 5G systems due to its increased spectral efficiency over conventional orthogonal multiple access (OMA). In NOMA, multiple users send data simultaneously in the uplink to the base station. The signals create mutual interference, but the base station can employ successive interference cancellation (SIC), i.e., decode one of the signals, and then subtract the corresponding codeword from the received signal, such that the other signal is interference-free. As a result, NOMA can increase the sum ergodic capacity of the system \cite{ding2014noma}. 

However, the ergodic capacity is not a meaningful performance metric for applications that require ultra-reliable low-latency communications (URLLC). For example, industrial control systems often require latencies of at most a few milliseconds. 
The probability of violating this deadline must be very small, with target values of $10^{-6}$ and below \cite{3gpp.22.804}.
In contrast to the ergodic sum capacity, the delay violation probability is affected by the SIC decoding order:
the user that is decoded first faces interference by the second user and thus experiences a lower data rate than if it were decoded last. 
Although the sum rate remains the same regardless of the decoding order, a very low rate for one of the users can mean that the user's data cannot be transmitted and must be buffered, leading to a queueing delay.
The delay performance of the two-user NOMA uplink thus depends on the optimal trade-off between the two decoding orders with respect to the users' delay constraints. 
\hlchange{Furthermore, both SIC decoding orders may lead to a low rate for one of the users. 
We need to consider a more general joint decoding scheme that can also achieve intermediate rate points. Even though joint decoding does not increase the ergodic sum capacity, it may improve the delay performance by avoiding very low rates.
}

The above discussion on the rate adaptation assumed that the base station has perfect knowledge of the SNR of both users, and that the users can communicate without errors at a rate equal to the capacity of the channel.
However, these assumptions become highly inaccurate for low-latency systems.
When the duration of each time slot is short, the overhead from estimating the instantaneous SNR affects the performance. More importantly, the training sequences used for estimation must be short, so that the base station will only have imperfect channel state information (CSI) for both users. The base station must then select the data rates based on that imperfect knowledge. 
The actual capacity of the channel may be smaller than the rate that was selected based on imperfect CSI, so that outages may occur. 
Furthermore, due to finite blocklength effects, decoding errors may even occur when the selected rate is lower than the actual capacity \cite{polyanskiy2010channel}.
In order to determine the delay performance of NOMA systems under these realistic assumptions, one must first determine the decoding error probabilities due to imperfect CSI and finite blocklength channel coding.

\subsection{Related Work}
This work combines an analysis of the queueing delay on the link layer of a wireless system with physical layer transmissions based on NOMA. The queueing delay of wireless systems in fading channels can be analyzed using frameworks such as effective capacity \cite{wu2003effective} or stochastic network calculus \cite{fidler2006network,alzubaidy2016ton}. While we consider in this work only stochastic network calculus, the resulting expressions can also be used to derive the effective capacity. We now discuss the state of the art on the delay performance of NOMA systems, as well as on the queueing performance of systems that are subject to imperfect CSI and finite blocklength coding.

\subsubsection{NOMA}
With respect to low-latency communications, several authors have considered the use of NOMA. 
Yang et al. \cite{yang2017optimality} optimize the sum rate subject to constraints on the minimum rates of each user.
Similarly, Timotheou et al. \cite{timotheou2015fairness} consider the max-min fairness for the individual rates, i.e., maximize the minimum rate among all users. However, the minimum rates may not be a meaningful metric for fading channels, where one of the rates may occasionally become very small.
In \cite{yang2016generalpower} and \cite{yang2017noma}, the outage probability in NOMA systems is analyzed. The outage probability is a meaningful metric for systems where the data rate is kept constant. However, not adapting the data rate to the channel state is generally suboptimal.

More specifically with respect to queueing analysis, several authors have considered NOMA.
Choi \cite{choi2017effective} studied the effective capacity of NOMA systems, assuming that one of the users is always decoded first (i.e., always suffers from interference).
Similarly, in our previous work \cite{schiessl2016interference}, we considered the impact of interference on the queueing performance, which corresponds to the performance of the user that is always decoded first.
The queuing performance of downlink NOMA systems was studied by Yu et al. \cite{yu2018link} (using effective capacity) and Xiao et al. \cite{xiao2019mimonoma} (using stochastic network calculus).
Unfortunately, even though both \cite{yu2018link} and \cite{xiao2019mimonoma} consider SIC decoding in the downlink, those results cannot be applied to SIC decoding the uplink. In the downlink, the signal for the weaker user must always be decoded under interference from the stronger user, whereas in the uplink, the decoding order can be varied. In the uplink, we also do not require the additional assumption that the first user always has higher instantaneous SNR than the second user (made in both \cite{yu2018link} and \cite{xiao2019mimonoma}), which does not always hold.
Close to our work is the work by Qiao et al.\cite{qiao2012transmission}, where the decoding order of the users in the uplink was varied based on the instantaneous channel states and on the individual QoS parameters. However, the authors assumed CSI to be perfect, and only considered SIC decoding, which may yield suboptimal rate points compared to a more general joint decoder. 

\subsubsection{Imperfect CSI and Finite-Length Coding}
Polyanskiy et. al \cite{polyanskiy2010channel} have studied bounds on the decoding error probability of channel codes with finite blocklength and also presented a closed-form normal approximation. Yang et. al \cite{yang2014quasi} extended these results to quasi-static block-fading channels. Scarlett et al. \cite{scarlett2017dispersion} studied finite blocklength effects in a multiuser scenario, where multiple users simultaneously communicate with finite blocklength channel codes. MolavianJazi \cite{molavianjazi2014phdthesis} studied finite blocklength effects in the two-user NOMA scenario. 

With respect to the delay performance, the effects of finite blocklength coding were studied in \cite{gursoy2013throughput,schiessl2015delay}.
In \cite{schiessl2016imperfectcsi}, we have studied the joint impact of imperfect CSI and finite blocklength effects on the delay performance of wireless systems, assuming a single-user single-antenna scenario. Furthermore, we studied the impact of these effects in a more general multiuser multi-antenna downlink scenario in \cite{schiessl2019miso}, where beamforming was applied to avoid interference.

\subsection{Contributions}

In this work, we apply the framework of stochastic network calculus (SNC) \cite{fidler2006network,alzubaidy2016ton} to study the performance of two-user NOMA uplink when both users must communicate within a short maximum delay.
Our contributions can be summarized as follows:
\begin{itemize}
	\item For perfect CSI and SIC decoding, we identify the problem of optimal rate selection (i.e., optimal decoding order) for quantized SNR distributions as a 0-1 knapsack problem that can be solved with a greedy algorithm.
	\item For perfect CSI and a more general joint decoder, we determine the optimal rate adaptation function in closed form.
	\item For imperfect CSI and SIC decoding, we derive closed-form approximations for the decoding error probabilities. This allows us to determine optimal rate allocations for this case. Simulation results show that the approximations are sufficiently accurate.
	\item Using methods from prior work, we present closed-form approximations for the decoding error probabilities under imperfect CSI and finite blocklength coding.
	\item Our numerical study shows that under ideal assumptions with perfect CSI and under tight delay constraints, NOMA with joint decoding significantly outperforms OMA when there is a large difference between the two users' average channels. NOMA with SIC decoding performs significantly worse, and may be worse than OMA, depending on the parameters.
	\item \hlchange{Imperfect CSI and finite blocklength effects cause a significant performance loss. Our results indicate that the performance loss is slightly larger for the NOMA than for OMA,}
\end{itemize}

This paper is structured as follows: For the ideal model with perfect CSI and very long codewords, we present the system model in Sec.~\ref{sec:system_model} and the delay analysis and rate optimization in Sec.~\ref{sec:analysis}. We then analyze imperfect CSI in Sec.~\ref{sec:icsi}, and finite blocklength coding in Sec.~\ref{sec:icsi_fbl}. Numerical results are given in Sec.~\ref{sec:numerics}. We present our conclusions in Sec.~\ref{sec:conclusions}.

%% file: System_Model.tex

\section{System Model}
\label{sec:system_model}
We analyze data transmissions in a multiple-access channel (MAC) where two devices send data packets to a central base station in a time-slotted fashion. We consider applications that generate periodic and time-critical data at the device/user side, which should be transmitted to the base station within a short deadline of $w$ time slots with high reliability.
In Sec.~\ref{ssec:system_channel}, we discuss the data transmission on the physical layer, where we consider only an ideal model with perfect CSI and infinitely long channel codes. Later, in Sec.~\ref{sec:icsi} and \ref{sec:icsi_fbl}, we will consider more realistic physical layer models with imperfect channel estimation and finite blocklength coding. 
Due to time-varying data rates and transmission errors at the physical layer, the devices must keep their data in a buffer for transmission in subsequent time slots. The resulting queueing delay is described in Sec.~\ref{ssec:system_queueing_model}.
We conclude this section with the problem statement in Sec.~\ref{ssec:system_problem}.

\subsection{Physical Layer Model}
\label{ssec:system_channel}
The channel is assumed to be block-fading, i.e., remains constant for the duration of one block or time slot of $\ntotal$ channel uses, and changes independently between time slots, which is an accurate model for example for systems that employ frequency hopping. We now consider a single time slot. For each channel use, the received signal $y$ is denoted as
\begin{equation}
y = h_1 x_1 + h_2 x_2 + z
\end{equation}
where $z$ is additive white Gaussian noise. Without loss of generality, we assume that $z\sim\mathcal{CN}(0,1)$ (unit variance) and that $\expected{|h_k|^2}=1$, such that the average power of the code symbols $\expected{|x_k|^2}=\Pk$ corresponds to the \emph{average} SNR of the signal at the receiver in case there is no interference. 
The instantaneous signal-to-noise ratio (SNR) of the received signal of user $k$ is denoted as $\snr_k=\snravg_k|h_k|^2$.
We assume Rayleigh-fading with $h_k\sim\mathcal{CN}(0,1)$.
For the initial analysis, we assume that the instantaneous SNR values $\snr_k$ are perfectly known at transmitting devices and at the base station, so that the entire time slot can be used for transmitting codewords $\xvec_k$ of length $\nd=\ntotal$ (for imperfect CSI, see Sec.~\ref{sec:icsi}). Furthermore, we assume that $\nd$ is sufficiently large so that error-free communication at a rate equal to the capacity is possible, using Gaussian codewords $\xvec_k$ (we discuss finite-length coding in Sec.~\ref{sec:icsi_fbl}).
The base station can try to decode the signals through successive interference cancellation (SIC) or jointly. For SIC decoding, assume that codeword $x_1$ is decoded first. This is possible if
\begin{align}
\rate_1 &< \RImin(\snr_1,\snr_2) \defined \log_2\left(1+\frac{\snr_1}{\snr_2+1}\right)
\label{eq:rate1min}
\,.
\end{align}
After successfully decoding the signal sent by user 1, the base station reconstructs the codeword $x_1$ and subtracts $h_1 x_1$ from the received signal $y$. Then, signal 2 can be decoded if 
\begin{align}
\rate_2 &< \RIImax(\snr_2) \defined \log_2\left(1+\snr_2\right)
\label{eq:rate2max}
\,.
\end{align}
The decoding order can also be reversed, such that $x_2$ is decoded first and then subtracted. 

\begin{figure}[t]
	\centering
	\def\svgwidth{8cm}
	\scriptsize{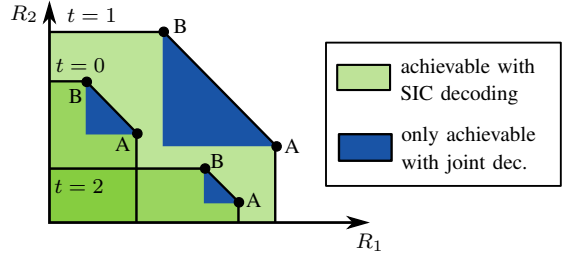}
	\caption{Schematic illustration of 3 random samples of the capacity region and achievable rate pairs for NOMA.}
	\label{fig:2user_mac_region}
\end{figure}

By using a decoder that decodes $x_1$ and $x_2$ jointly, the base station can decode both $x_1$ and $x_2$ whenever the rates $\rate_1$ and $\rate_2$ are inside the capacity region, which is given as \cite{tse2005fundamentals}:
\begin{align}
\rate_1 &< \RImax(\snr_1) = \log_2(1+\snr_1)
\label{eq:rate1}\\
\rate_2 &< \RIImax(\snr_2) = \log_2(1+\snr_2)
\label{eq:rate2}\\
\rate_1+\rate_2 &< \Rsum(\snr_1,\snr_2) \defined \log_2(1+\snr_1+\snr_2)
\,.
\label{eq:sumrate}
\end{align}
The capacity regions for three time slots, i.e., three random instances of $\snr_1$ and $\snr_2$, are illustrated in Fig.~\ref{fig:2user_mac_region}. The capacity region always has the shape of a pentagon, with two of the corner points \hlchangeb{(denoted as A and B)} corresponding to the \hlchangeb{maximum} rates achieved by the SIC decoder.\footnote{\hlchangeb{The rates in \eqref{eq:rate1min} to \eqref{eq:sumrate} can be chosen arbitrarily close to the capacity or capacity region. In order to simplify discussions, we assume that \eqref{eq:rate1min} to \eqref{eq:sumrate} also hold when the rates are equal to the capacity or on the boundary of the capacity region.}} \hlchangeb{With} joint decoding\hlchangeb{, one} can also achieve the rate pairs on the segment $\overline{\mathrm{AB}}$ between the two corner points.\footnote{The points on the segment between the two SIC points can also be achieved through SIC with time sharing between the two decoding orders, or through a rate-splitting approach \cite{rimoldi1996ratesplitting}. However, for better clarity, we apply the term ``SIC" (``NOMA-SIC") exclusively to the case where only the corner points of the rate regions can be selected, and we apply the term ``joint decoding" (``NOMA-joint") exclusively to the case where all rates in the capacity region can be selected.} 
\hlchangeb{We assume for now that} in each time slot, the base station knows the instantaneous SNR values $(\snr_1,\snr_2)$ \hlchangeb{perfectly and selects an achievable rate pair, which is} denoted as $\rate_1=\rateadapt_1(\snr_1,\snr_2)$ and $\rate_2=\rateadapt_2(\snr_1,\snr_2)$, or $(\rate_1,\rate_2)=\rateadaptvec(\snr_1,\snr_2)$ for short.
\hlchangeb{There is no reason to select a rate pair that is below the maximally achievable rates, so the base station will select} 
a pair of rates $(\rate_1,\rate_2)$ that either corresponds to one of the corner points \hlchangeb{A/B} (in case of SIC decoding), or to any point \hlchangeb{on the segment $\overline{\mathrm{AB}}$} (in case of joint decoding). \hlchangeb{The selected rates are then signaled} to the users through a feedback link, which we assume to be instantaneous and error-free.

\subsection{Queueing Model}
\label{ssec:system_queueing_model}
Due to the time-varying channel conditions, the data arriving at the users must be stored in a buffer until successful transmission. This leads to a random queueing delay. A queueing system is described by its arrival, service, and departure processes. The arrival process $\Abit_k(\ts)$ describes the amount of data in bits that arrives and is put into the buffer at user $k$ in time slot $\ts$. The service process $\Sbit_k(\ts)$ depends on the instantaneous SNR values $\snr_1(\ts)$ and $\snr_2(\ts)$ in time slot $\ts$, which change from time slot to time slot. The rate is determined by the rate adaptation scheme as $\rate_k(\ts)=\rateadapt_k(\snr_1(\ts),\snr_2(\ts))$. In case the transmission is successful, the service is $\Sbit_k(\ts)=\nd \rate_{k}(\ts)$. In case of a transmission error (which will become relevant in Sec.~\ref{sec:icsi} and Sec.~\ref{sec:icsi_fbl}), the base station will indicate the error event through a feedback bit
so that the users will not remove the corresponding data from the queue, which corresponds to $\Sbit_k(\ts)=0$. 
The departure process $\Dbit_k(\ts)$ describes the data that is actually transmitted over the wireless channel, which is the minimum of $\Sbit_k(\ts)$ and the amount of data waiting in the buffer. 
The virtual delay $\Delay_k(\ts)$ of the data arriving in time slot $\ts$ at user $k$ is then defined as \cite{alzubaidy2016ton}
\begin{align}
\Delay_k(\ts) \defined \inf \left\{u\ge 0: \quad \sum_{i=0}^{\ts-1} \Abit_k(i) \le \sum_{i=0}^{\ts+u-1} \Dbit_k(i) \right\}
\label{eq:def_delay}
\end{align}
The delay $\Delay_k(\ts)$ is a random variable, and its distribution can be stated in terms of the delay violation probability over all time slots $\ts$ with respect to a certain deadline $w$:
\begin{align}
\pvk(w) \defined \sup_{t\ge 0} \, \Prob{\Delay_k(\ts) > w}
\label{eq:def_pdelayviol}
\end{align}

\subsection{Problem Statement}
\label{ssec:system_problem}
In this work, we study the delay performance of NOMA systems with SIC or joint decoding in order to determine whether NOMA is superior to OMA when operating under strict delay constraints, and whether joint decoding significantly improves the performance over SIC decoding. The delay performance of NOMA depends on the optimal rate allocation scheme $\rateadaptvec$. In order to ensure reliable low-latency communications for both users, we need to determine $\rateadaptvec$ such that the delay violation probabilities $\pvk(w)$ for both users are jointly minimized. 
We first consider the ideal model with perfect CSI, and find the optimal rate allocation for SIC decoding and for joint decoding. 
However, in more realistic models with imperfect CSI and finite blocklength codes, rate adaptation is more difficult, as we have to find an optimal trade-off between the selected rates and the corresponding decoding error probabilities.
We can then address the following questions: can NOMA outperform OMA under short delay constraints also with realistic system models? And how large is the difference between SIC decoding and joint decoding in that case?

\hlremove{Remove: In general, the problem cannot be solved analytically, as the delay violation probability $\pvk(w)$ cannot be determined analytically. However, $\pvk(w)$ can be approximated using the framework of effective capacity \cite{wu2003effective}, or bounded through stochastic network calculus (SNC) \cite{fidler2006network,alzubaidy2016ton}. SNC provides an analytical upper bound on the delay violation probability, i.e., it provides performance guarantees for reliable low-latency communications.
We will therefore use the expressions from SNC to solve the optimal rate adaptation problem analytically.}

%% file: figures/rate_region.pdf_tex
\begingroup%
  \makeatletter%
  \providecommand\color[2][]{%
    \errmessage{(Inkscape) Color is used for the text in Inkscape, but the package 'color.sty' is not loaded}%
    \renewcommand\color[2][]{}%
  }%
  \providecommand\transparent[1]{%
    \errmessage{(Inkscape) Transparency is used (non-zero) for the text in Inkscape, but the package 'transparent.sty' is not loaded}%
    \renewcommand\transparent[1]{}%
  }%
  \providecommand\rotatebox[2]{#2}%
  \ifx\svgwidth\undefined%
    \setlength{\unitlength}{307.21521333bp}%
    \ifx\svgscale\undefined%
      \relax%
    \else%
      \setlength{\unitlength}{\unitlength * \real{\svgscale}}%
    \fi%
  \else%
    \setlength{\unitlength}{\svgwidth}%
  \fi%
  \global\let\svgwidth\undefined%
  \global\let\svgscale\undefined%
  \makeatother%
  \begin{picture}(1,0.40400677)%
    \put(0,0){\includegraphics[width=\unitlength,page=1]{rate_region.pdf}}%
    \put(0.45451544,0.15032613){\color[rgb]{0,0,0}\makebox(0,0)[lb]{\smash{A}}}%
    \put(0.26977387,0.3491267){\color[rgb]{0,0,0}\makebox(0,0)[lb]{\smash{B}}}%
    \put(0.39033951,0.06014702){\color[rgb]{0,0,0}\makebox(0,0)[lb]{\smash{A}}}%
    \put(0.33140361,0.12595222){\color[rgb]{0,0,0}\makebox(0,0)[lb]{\smash{B}}}%
    \put(0.64851847,0.17070911){\color[rgb]{0,0,0}\makebox(0,0)[lb]{\smash{only achievable}}}%
    \put(0.64764241,0.28312379){\color[rgb]{0,0,0}\makebox(0,0)[lb]{\smash{achievable with}}}%
    \put(0.6489635,0.12454435){\color[rgb]{0,0,0}\makebox(0,0)[lb]{\smash{with joint dec.}}}%
    \put(0.64697487,0.23778613){\color[rgb]{0,0,0}\makebox(0,0)[lb]{\smash{SIC decoding}}}%
    \put(0.07434779,0.28498829){\color[rgb]{0,0,0}\makebox(0,0)[lb]{\smash{$t=0$}}}%
    \put(0.07199193,0.08446642){\color[rgb]{0,0,0}\makebox(0,0)[lb]{\smash{$t=2$}}}%
    \put(0.09330057,0.36875855){\color[rgb]{0,0,0}\makebox(0,0)[lb]{\smash{$t=1$}}}%
    \put(0.5727625,-0.00711611){\color[rgb]{0,0,0}\makebox(0,0)[lb]{\smash{$R_1$}}}%
    \put(-0.00147812,0.37231796){\color[rgb]{0,0,0}\makebox(0,0)[lb]{\smash{$R_2$}}}%
    \put(0.09565341,0.23361735){\color[rgb]{0,0,0}\makebox(0,0)[lb]{\smash{B}}}%
    \put(0.17316072,0.15437809){\color[rgb]{0,0,0}\makebox(0,0)[lb]{\smash{A}}}%
  \end{picture}%
\endgroup%

%% file: Analysis.tex

\section{Analysis -- Ideal Case}
\label{sec:analysis}
In this section, we will first present in Sec.~\ref{ssec:queueing} the analytical upper bound on the delay violation probability from stochastic network calculus. Then, we show in Sec.~\ref{ssec:general_rate_alloc} how this result can be used to reformulate the problem statement analytically as an optimization problem.
We will then solve the optimal rate adaptation problem for SIC decoding and joint decoding in Sec.~\ref{ssec:pcsi_rates_sic} and Sec.~\ref{ssec:pcsi_rates_joint}, respectively. In order to introduce the basic optimization methodology, we consider in this section only the ideal system model from Sec.~\ref{ssec:system_channel}. However, we are eventually interested in the delay performance for a more realistic system model with imperfect CSI and finite-length coding, which we will introduce in Sec.~\ref{sec:icsi} and \ref{sec:icsi_fbl}.

\subsection{Stochastic Network Calculus}
\label{ssec:queueing}
We now give a brief summary of previous results from stochastic network calculus (SNC). Specifically, we show an upper bound from SNC on the delay violation probability $\pvk(w)$ in \eqref{eq:def_pdelayviol} \cite{fidler2006network,alzubaidy2016ton}. 
This summary closely follows the summary given in our previous works \cite{schiessl2015delay,schiessl2016imperfectcsi,schiessl2019miso}.

We follow \cite{alzubaidy2016ton}, where SNC is applied in a transform domain, also referred to as \emph{SNR-domain}. The bit-domain arrival and service processes $\Abit_k(\ts)$ and $\Sbit_k(\ts)$ defined in Sec.~\ref{ssec:system_queueing_model} are transformed to the SNR-domain via the exponential function: $\Asnr_k(\ts) \defined e^{\Abit_k(\ts)}$ and $\Ssnr_k(\ts) \defined e^{\Sbit_k(\ts)}$.
An upper bound on the delay violation probability $\pvk(w)$ can then be computed in terms of the Mellin transforms of $\Asnr_k(\ts)$ and $\Ssnr_k(\ts)$, where we can omit the time index $\ts$ because of i.i.d. arrivals and block-fading.
The Mellin transform of a nonnegative random variable $\Xsnr$ is defined as \cite{alzubaidy2016ton}
\begin{equation}
\Mellin_\Xsnr(\s)\defined\mathbb{E}\left[\Xsnr^{\s-1}\right]
\end{equation}
for a parameter $\s \in \mathbb{R}$. 
For the analysis, we always choose $\s_k>0$ and first check whether the stability condition $\Mellin_{\Asnr_k}(1+\s_k)\Mellin_{\Ssnr_k}(1-\s_k) < 1$ holds.
If it holds, define the kernel \cite{alzubaidy2016ton,schiessl2015delay}
\begin{align}
\Kernelk{k}{\s_k,w} &\defined 
 \frac{\Mellin_{\Ssnr_k}(1-\s_k)^{w}}{1-\Mellin_{\Asnr_k}(1+\s_k)\Mellin_{\Ssnr_k}(1-\s_k)} 
\;.
\label{eq:snc_kernel}
\end{align}
This kernel is strictly monotonically increasing in both $\Mellin_{\Asnr_k}(1+\s_k)$ and $\Mellin_{\Ssnr_k}(1-\s_k)$, and provides an upper bound for the delay violation probability, which holds for any time slot $t$, including the limit $t\to\infty$ (steady-state):
\begin{equation}
\pvk(w) \leq \inf_{\s_k>0}\left\{ \Kernelk{k}{\s_k,w} \right\} \;.
\label{eq:pdelay_bound}
\end{equation}
This bound holds for any the parameter $\s_k>0$. In order to find the tightest upper bound on $\pvk(w)$, one should iterate over $\s_k$.

\subsection{Rate Allocation Problem}
\label{ssec:general_rate_alloc}
We seek to determine a rate allocation scheme $\rateadaptvec$ which jointly minimizes the delay violation probabilities $\pvk(w)$ in \eqref{eq:def_pdelayviol} of the two users $k\in\{1,2\}$. In order to work with analytical expressions of the system, we use the analytical upper bound \eqref{eq:pdelay_bound} on $\pvk(w)$ based on the kernels $\Kernelk{k}{\s_k,w}$.
Then, the optimization of the rate allocation scheme $\rateadaptvec$ can be formulated as follows: given specific QoS constraints for the first user, how should the base station select the rates such that the bound on $\pvk(w)$ for the second user is minimized? This is written as
\begin{align}
\begin{aligned}
	\argmin_{\rateadaptvec} \quad & \inf_{\s_2>0}\left\{ \Kernelk{2}{\s_2,w} \right\}
	\\ \text{s.t.} \quad & \inf_{\s_1>0}\left\{ \Kernelk{1}{\s_1,w} \right\}\leq \pvmax
	\end{aligned}
	\tag{P.I}\label{opt_rates_kernel}
\end{align}
The kernels depend on the rate adaptation function $\rateadaptvec$ through the Mellin transform of the SNR-domain service process $\Mellin_{\Ssnr_k}(1-\s_k) = \expected{e^{-\s_k\ntotal\rateadapt_k(\snrvec)}}$.
We iterate over all possible combinations of $\s_k$ and solve the problem for a specific choice of the values $\s_k$. By inspecting \eqref{eq:snc_kernel}, we can deduce that the kernel $\mathcal{K}_k\left(\s_k,w\right)$ is monotonically increasing in $\Mellin_{\Ssnr_k}(1-\s_k)$, as long as the stability condition $\Mellin_{\Asnr_k}(1+\s_k)\Mellin_{\Ssnr_k}(1-\s_k) < 1$ is satisfied. 
Therefore, the optimization problem can be formulated directly in terms of $\Mellin_{\Ssnr_k}(1-\s_k)$ instead of $\Kernelk{k}{\s_k,w}$:
\begin{align}
\begin{aligned}
\argmin_{\rateadaptvec} \quad & \Mellin_{\Ssnr_2}(1-\s_2)
\\\text{s.t.} \quad & \Mellin_{\Ssnr_1}(1-\s_1) \leq \pvmaxtilde
\end{aligned}
\tag{P.II}\label{opt-PI}
\end{align}
with $\pvmaxtilde$ chosen such that $\Kernelk{1}{\s_1,w} \leq \pvmax$.

\subsection{Rate Allocation for SIC}
In case the base station employs SIC decoding, the rate allocation can be found as follows:
\label{ssec:pcsi_rates_sic}
\begin{result}
	\label{result_rates_pcsi_sic}
	When using SIC decoding, i.e., when the rate scheduler can only decide between the two rate pairs $(\RImax,\RIImin)$ or $(\RImin,\RIImax)$, the optimal solution to problem \eqref{opt-PI}
	is given by
	\begin{align}
	\rateadapt_1(\snr_1,\snr_2) &=  \left\{
	\begin{array}{ll}
	\RImin(\snr_1,\snr_2) &\quad\quad\text{for }\VWR(\snr_1,\snr_2)>\lambda \\
	 \RImax(\snr_1) &\quad\quad\text{otherwise}\\
	 \end{array}
	 \right.
	 \\
	 \rateadapt_2(\snr_1,\snr_2) &= \Rsum(\snr_1,\snr_2)-\rateadapt_1(\snr_1,\snr_2)
	\,
	\end{align}
	with value-to-weight ratio
	\begin{align}
	\VWR(\snr_1,\snr_2) = \frac{e^{-\s_2\ntotal\RIImin(\snr_1,\snr_2)}-e^{-\s_2\ntotal\RIImax(\snr_2)}}{e^{-\s_1\ntotal\RImin(\snr_1,\snr_2)}-e^{-\s_1\ntotal\RImax(\snr_1)}}
	\end{align}
	and $\lambda>0$ is the smallest value such that $\Mellin_{\Ssnr_1}(1-\s_1) \leq \pvmaxtilde$ is still satisfied.
\end{result}
\begin{proof}
Without loss of generality, we assume $\ntotal=1$.
We discretize the distributions of $\snr_1$ and $\snr_2$ to points $i=1\ldots \Npoints$ and rewrite the optimization problem in terms of $x_{i}\in\{0,1\}$, where $x_{i}=0$ means that decoding order A is selected ($\rate_{1,i}=\RImaxi$ and $\rate_{2,i}=\RIImini$), and $x_{i}=1$ means that point B is selected ($\rate_{1,i}=\RImini$ and $\rate_{2,i}=\RIImaxi$):
\begin{equation}
\begin{aligned}
\argmin_{ x_{i} \in\{0,1\} } \quad & \sum_i p_{i}e^{-\s_2(\RIImini+x_i\left(\RIImaxi-\RIImini\right))}
\\\text{s.t.} \quad & \sum_i p_{i}e^{-\s_1 \left(\RImaxi+x_i\left(\RImini-\RImaxi\right)\right)} \leq \pvmaxtilde
\end{aligned}
\tag{P.IIa}\label{opt-PIa}
\end{equation}
For $x_{i} \in\{0,1\}$, the problem can be converted into the following equivalent problem:
\begin{equation}
\begin{aligned}
\argmax_{ x_{i} \in\{0,1\} }  \quad & z=\sum_i x_i v_i
\\\text{s.t.} \quad & \sum_i x_i w_i\leq \pvmaxtildealt
\end{aligned}
\tag{P.IIb}\label{opt-PIb}
\end{equation}
with
\begin{align}
v_i &= p_i\left( e^{-\s_2\RIImini}-e^{-\s_2\RIImaxi}\right)
\\
w_i&= p_i\left(e^{-\s_1\RImini}-e^{-\s_1 \RImaxi}\right)
\\
\pvmaxtildealt&=\pvmaxtilde-\sum_i p_i e^{-\s_1\RImaxi}
\end{align}
We identify this optimization problem as the well-known 0-1 knapsack problem of selecting items with value $v_i \geq 0$ and weight $w_i \geq 0$ subject to a weight limit $\pvmaxtildealt$. 
We assume $0 < \pvmaxtildealt <\sum_i w_i$, otherwise the problem is either trivial or infeasible. 
To solve this problem, we first allow $x_{i}$ in \eqref{opt-PIb} to vary continuously from 0 to 1, i.e., we relax the problem \cite{freville2004multidimensional}.\footnote{The problem \eqref{opt-PIb} is only equivalent to \eqref{opt-PIa} when $x_k$ is integer, and relaxing \eqref{opt-PIb} is thus \emph{not} equivalent to relaxing the integer constraint in \eqref{opt-PIa}, i.e., not equivalent to allowing rates between the extreme points A and B. 
}
The continuous problem ($\relaxedprob$) can be easily solved using a greedy algorithm, and provides an approximate solution to the discrete problem. 
Specifically, Dantzig \cite{dantzig1957discrete} showed that the optimal solution to $\relaxedprob$ can be found by ordering the items decreasingly according to their value-to-weight ratios $\VWR_i=v_i/w_i$ and then selecting the first $j-1$ items with the highest value-to-weight ratios ($x_i=1$ for $i=1,\ldots,j-1$, after reordering) such that $\sum_{i=1}^{j-1} x_i w_i\leq \pvmaxtildealt$ is still satisfied. Then, for the $j$-th item, only a fractional value of $0\leq x_j<1$ is chosen, such that $\sum_{i=1}^{j} x_i w_i$ becomes equal to $\pvmaxtilde$. The remaining items are not selected.
Clearly, the optimal value $z^*$ of problem \eqref{opt-PIb} cannot exceed the optimal value $z^*_\relaxedprob$ of the continuous problem: $z^*\leq z^*_\relaxedprob$. Furthermore, we obtain a rounded solution by setting $x_j=0$, with a corresponding value of $z^*_{\lfloor\relaxedprob\rfloor}$. The rounded solution is a possible solution to \eqref{opt-PIb}, thus $z^*_{\lfloor\relaxedprob\rfloor}\leq z^*$. As $z^*_\relaxedprob-z^*_{\lfloor\relaxedprob\rfloor}=x_j v_j$, with $v_j$ vanishing as the quantization intervals and the corresponding probability masses $p_i$ tend to zero, the output of the greedy algorithm converges to the optimal solution.
\end{proof}

\hlremove{
\begin{remark}
	\hlchange{
		Even though this paper only considers Rayleigh fading channels, Result~\ref{result_rates_pcsi_sic} as well as the following Result~\ref{result_rates_pcsi_joint} also apply to other fading distributions.}
\end{remark}
}
\begin{remark}
\hlchange{The problem above was already addressed in \cite{qiao2012transmission}, which provided a function that denotes the boundary of the two regions where user 1 or user 2 is decoded first, respectively. 
However, the proof relies on a specific result from variational calculus, which requires that the end points of the function are fixed and known, which may not be the case.}
\end{remark}

\subsection{Rate Allocation for Joint Decoding}
With SIC decoding, only the rate pairs on the two corner points of the capacity region are achievable. On the other hand, joint decoding allows rate points in between the corner points.
\label{ssec:pcsi_rates_joint}
\begin{result}
	\label{result_rates_pcsi_joint}
Under joint decoding, i.e., when all rates in the achievable rate region can be selected, the optimal solution to the rate adaptation problem \eqref{opt-PI}
is given by
\begin{align}
\rateadapt_1(\snrvec) &= \left\{\begin{array}{ll}
\RImin(\snr_1,\snr_2) & \textnormal{if } 
\RIhat(\snrvec) < \RImin(\snr_1,\snr_2)  \\
\RImax(\snr_1) & \textnormal{if }\RIhat(\snrvec) > \RImax(\snr_1) \\
\RIhat(\snrvec) & \textnormal{otherwise}\\
\end{array}
\right.
\label{eq:R1hat_cont}
\\ \rateadapt_2(\snrvec) &= \Rsum(\snrvec) - \rateadapt_1(\snrvec)
\,,
\end{align}
where $\RIhat(\snrvec) = \frac{\s_2}{\s_1+\s_2}\Rsum(\snrvec)+\tilde{\lambda}_1
\label{eq:R1hat}$ and
$\tilde{\lambda}_1 \in \mathbb{R}$ is the smallest value such that $\Mellin_{\Ssnr_1}(1-\s_1) \leq \pvmaxtilde$ is still satisfied. 
\hlremove{Furthermore, the three decision regions for $\rateadapt_1(\snrvec)$ as identified by \eqref{eq:R1hat_cont} are homogeneous and are limited by functions $g_1(\snr_1)$ and $g_2(\snr_2)$.}
\end{result}
\begin{proof}
Without loss of generality, assume $\ntotal=1$.
We quantize the joint distribution of $\snr_1$ and $\snr_2$ to points labeled as $i=1\ldots \Npoints$ with probability mass $p_i$.
Note that if the sum rate constraint \eqref{eq:sumrate} does not hold with equality, then one of the rates could be increased without penalty for the other user. An optimal rate allocation will thus always select a rate pair which satisfies the sum rate constraint. We then have $\RIIi=\Rsumi-\RIi$ and rewrite the optimization problem:
\begin{align}
\argmin_{ \RIi } \quad & 
\sum_i\ p_{i}e^{-\s_2(\Rsumi- \RIi)}
\\\text{s.t.} \quad & \sum_i p_{i}e^{-\s_1 \RIi} \leq \pvmaxtilde
\label{eq:constraint1}
\\& \RIi \leq \log_2(1+\snr_{1,i}) & i = 1,\ldots,\Npoints
\label{eq:constraint2}
\\&  \RIi \geq \log_2\left(1+\frac{\snr_{1,i}}{\snr_{2,i}+1}\right)  & i = 1,\ldots,\Npoints
\label{eq:constraint3}
\end{align}
The problem is convex in $\RIi$. We associate the Lagrange multipliers $\lambda_1\geq 0$, $\mu_{i}\geq 0$ and $\nu_{i}\geq 0$ with the contraints (\ref{eq:constraint1}), (\ref{eq:constraint2}), and (\ref{eq:constraint3}), respectively.
According to the Karush-Kuhn-Tucker conditions \cite{boyd2004convex}, the optimal rates $\RIi$ for all $i=1,\ldots,\Npoints$ must satisfy
\begin{align}
\s_2e^{-\s_2(\Rsumi -\RIi)} -\lambda_1 \s_1e^{-\s_1 \RIi} +\mu_i - \nu_i = 0
\,.
\label{eq:gradient_lagrange_2user}
\end{align}
When either of the contraints \eqref{eq:constraint2} and \eqref{eq:constraint3} is satisfied with equality, then the value of $\RIi$ is known. Otherwise, there is slackness in the constraints \eqref{eq:constraint2} and \eqref{eq:constraint3}, and the complementary slackness conditions \cite{boyd2004convex} mandate that $\mu_i$ and $\nu_i$ must be zero. It follows for those cases:
\begin{align}
\lambda_1 \frac{\s_1}{\s_2}e^{-\s_1 \RIi} &= e^{-\s_2\left(\Rsumi- \RIi\right)}
\,.
\end{align}
With the definition $\tilde{\lambda}_1=\frac{1}{\s_1+\s_2} \log\left(\lambda_1\frac{\s_1}{\s_2}\right)$, we derive
\begin{align}
\RIi &= 
\frac{\s_2}{\s_1+\s_2}\Rsumi+\tilde{\lambda}_1 \defined \RIhat(\snrveci)
\label{eq:R1i}
\end{align}
For a given $\tilde{\lambda}_1$, the rates can be found by first computing $\RIhat(\snrveci)$ according to \eqref{eq:R1i}. For those SNR points $i$ where \eqref{eq:R1i} would violate the constraints \eqref{eq:constraint2} or \eqref{eq:constraint3},
the respective constraint must hold with equality.
\end{proof}

%% file: imp_csi.tex
\section{Imperfect CSI}
\label{sec:icsi}
\subsection{System Model and Approximations for Imperfect CSI}
In a realistic system, the users do not have perfect knowledge of the channel states. Instead, the channels must be estimated first, and the rates must be chosen based on the imperfect channel estimate. We assume that the two users send known training sequences of length $\ntk{1}$ and $\ntk{2}$ at the beginning of each time slot. The two users send their training sequences orthogonally (i.e., without interference) to the receiver, which obtains the MMSE channel estimates $\hmmse_k$, $k\in\{1,2\}$. The actual channel coefficients $H_k$ are unknown, i.e., random. 
\hlremove{
In the rest of the paper, we assume Rayleigh fading, i.e., $H_k\sim\mathcal{CN}(0,1)$. 
}
Given the MMSE estimates $\hmmse_k$, the actual channel coefficients are given as \cite{caire2010multiuser,schiessl2016imperfectcsi}
\begin{equation}
H_k=\hmmse_k+\Herr_k
\,,
\end{equation}
with $\Herr_k\sim\mathcal{CN}(0,\sigmazk^2)$,
\begin{equation}
\sigmazk^2=\frac{1}{1+\Ptk\ntk{k}}
\,,
\end{equation}
and $\Ptk$ denoting the SNR during the training phase of user $k$.
The actual SNR $\Snr_k=\snravg_k|H_k|^2$ of signal $k$ is then given as \cite{schiessl2016imperfectcsi}:
\begin{equation}
\Snr_k = \snravg_k|\hmmse_k|^2 + 2\snravg_k|\hmmse_k|\Re\left\{e^{-j\angle(\hmmse_k)}\Herr_k \right\} + \snravg_k|\Herr_k|^2
\label{eq:snr_snrmmse_equation}
\,.
\end{equation}
With sufficient training, the estimation error is relatively small, i.e., $|\Herr_k|\ll |\hmmse_k|$, and the last term becomes negligible. 
The term $\Re\{e^{-j\angle(\hmmse_k)}\Herr_k \}$ is Gaussian distributed with variance $\sigma_{Z,k}^2/2$, so we can approximate the distribution of the SNR $\Snr_k$ of signal $k$ as \cite{schiessl2016imperfectcsi}
\begin{equation}
\Snr_k \sim\mathcal{N}(\snrmmse_k, \sigmaicsik^2)
\label{eq:snr_gauss_approx}
\end{equation}
with $\snrmmse_k=\snravg_k|\hmmse_k|^2$, and
\begin{equation}
\sigmaicsik^2 = 2\snravg_k^2 |\hmmse_k|^2 \sigmazk^2 = 2\snravg_k \snrmmse_k \sigmazk^2
\,.
\label{eq:sigmaicsi}
\end{equation}

After the channel estimation, the base station must select the pair of rates $(\rate_1,\rate_2)$ at which the users should encode their data, based on the channel estimates $\snrmmse_1$ and $\snrmmse_2$. The users are informed about the chosen rates through an error-free feedback channel with zero delay.\footnote{For the single user case, we found that quantizing the rate to 6 bits is sufficient \cite{schiessl2016imperfectcsi}. It is reasonable to assume that the base station, which can operate at high transmit power, can communicate this small amount of data to the users with very low error probabilities and delays that are negligible compared to those in the uplink transmissions.}
While the rates are chosen based on the estimated SNR, i.e., $(\rate_1,\rate_2)=\rateadaptvec(\snrmmse_1,\snrmmse_2)$, the actual SNR of the channels, $\Snr_1$ and $\Snr_2$, is unknown and may be too low, such that decoding of the data sent by the users will fail.
The corresponding error probabilities depend on the chosen rates, as well as on the type of decoder used at the base station. 
In Sec.~\ref{ssec:icsi_cornerpoints}, we will derive and approximate the error probabilities for a simple SIC decoder, and in Sec.~\ref{ssec:icsi_joint}, we consider errors under joint decoding. We assume for now that the blocklength $\nd$ is very large. In Sec.~\ref{sec:icsi_fbl}, we will also consider the effects of finite blocklength channel coding, i.e., the case where $\nd$ is small.
It is assumed throughout the paper that rate adaptation is based on imperfect CSI, but that the receiver has perfect CSI when decoding the signal.\footnote{Additional pilot symbols sent during the data transmission phase will make the CSI at the receiver almost perfect compared to the imperfect estimate during rate selection. The receiver may also employ joint estimation and decoding \cite{skoglund2002code}. In \cite{schiessl2016imperfectcsi}, we computed an achievability bound for finite-length codes, which showed that after imperfect CSI during rate selection is taken into account, the additional performance impact due to imperfect CSI at the receiver is small.}
We note that when $\ntk{1}$ plus $\ntk{2}$ symbols are required for channel estimation in each time slot, only
$
\nd=\ntotal-\ntk{1}-\ntk{2}
$
symbols remain for the data transmission.

\subsection{Rate Optimization}
As discussed in Sec.~\ref{ssec:general_rate_alloc}, the optimal rate allocation problem can be formulated as minimizing the Mellin transform $\Mellin_{\Ssnr_2}(1-\s_2)$ of the SNR-domain service process for the second user, subject to a constraint on $\Mellin_{\Ssnr_1}(1-\s_1)$ for the first user.
When decoding errors occur with probability $\varepsilon_{k}$, the Mellin transform of the service process is given as \cite{schiessl2016imperfectcsi}
\begin{align}
\Mellin_{\Ssnr_k}(1-\s_k) = \expected{\varepsilon_{k}+(1-\varepsilon_{k})e^{-\theta_k \rate_{k}}}
\end{align}
For the original rate adaptation problem under perfect CSI, it is clear that $\Mellin_{\Ssnr_k}(1-\s_k)$ is convex in the chosen rates. We conjecture that the rate adaptation problem remains convex under imperfect CSI, which is also motivated by the findings for the single-user scenario in \cite{schiessl2016imperfectcsi}. In this paper, we will solve the problem for different values of $\pvmaxtilde$ in order to show the full range of possible trade-offs between user 1 and user 2. We obtain the same solutions if we consider the Lagrangian dual problem
\begin{align}
\argmin_{\rateadapt} \quad &\Mellin_{\Ssnr_2}(1-\s_2) + \tilde{\lambda} \Mellin_{\Ssnr_1}(1-\s_1)
\end{align}
and iterate over $\tilde{\lambda}$ instead of $\tilde{c}$.
When quantizing the joint distribution of $\Snr_1$ and $\Snr_2$ to $\Npoints$ points, each point $i$ having probability $p_i$, the problem becomes
\begin{equation}
\argmin_{ \rate_{1,i},\rate_{2,i} }  \quad \sum_{i=1}^{\Npoints} p_i\left( \varepsilon_{2,i}+(1-\varepsilon_{2,i})e^{-\theta_2 \rate_{2,i}}+
\tilde{\lambda}\left(\varepsilon_{1,i} + (1-\varepsilon_{1,i})e^{-\theta_1 \rate_{1,i}}\right)
\right)
\label{eq:argmin_rates_discrete}
\end{equation}
where $\varepsilon_{1,i}$ and $\varepsilon_{2,i}$ depend on both $\rate_{1,i}$ and $\rate_{2,i}$.
The optimal $\rateadaptvec\opt$ can be determined by finding the optimal rate pairs $(\rate_{1,i},\rate_{2,i})$ individually for each quantized point $i$ of the SNR distribution. However, to solve the problem, the error probabilities $\varepsilon_{1,i}$ and $\varepsilon_{2,i}$ must be computed for each quantized point $i$ and each considered rate pair. 
\hlchange{Assume that the SNR distribution is quantized on a grid with $N\times N$ points, and that for each of those points, we consider $M \times M$ different rate pairs. If one would need to perform a numerical, 2-dimensional integration to determine the error probabilities for each of the $N^2M^2$ rate/SNR combinations, then the rate adaptation problem would quickly become computationally infeasible, even with a coarse quantization of the distribution and the rate points.
}
Thus, in the following sections, we will present closed-form approximations for $\varepsilon_{1,i}$ and $\varepsilon_{2,i}$, which can be efficiently computed.

\subsection{Outage Probability under SIC Decoding}
\label{ssec:icsi_cornerpoints}
We first focus on the case where SIC decoding is used.
In the ideal model, the scheduler would select a rate pair $(\rate_1,\rate_2)$ that corresponds to one of the two corner points of the capacity region. However, in case of imperfect CSI, the transmitter only knows the estimated SNR $\snrmmse_1$ and $\snrmmse_2$. 
\hlchangeb{In the event that the selected rate pair $(\rate_1,\rate_2)$ lies outside the actual capacity region (defined by $\Snr_1$ and $\Snr_2$, whose exact values are unknown), a decoding error occurs.}
In order to ensure that this event occurs only with small probability, the scheduler \hlchangeb{should not select the rates exactly at the corner points of the estimated capacity region} but should select smaller rates.
Assume for now that the rates are close to the corner point where user 1 has priority in terms of rate, i.e., the base station will try to decode user 2 first such that user 1 experiences no interference.\footnote{Due to imperfect CSI, there is now a chance that the channel conditions are reversed, such that user 1 could be decoded first. We still obtain an approximate upper bound on the error probability by ignoring these highly unlikely events.}
The receiver can decode signal 2 directly if $\rate_2\le \log_2\left(1+\Snr_2/(\Snr_1+1)\right)$.
Then, signal 1 can be decoded if the receiver can decode and remove signal 2, and if at the same time $ \rate_1 \le \log_2\left(1+\Snr_1\right)$.
We define $\snrfactor_k=2^{\rate_k}-1$. The probability that user 1 cannot be decoded is then
\begin{align}
\epsone = \Prob{\Snr_1 < \snrfactor_1 \quad \hlchangeb{\vee}
	\quad  \frac{\Snr_2}{\Snr_1+1} < \snrfactor_2} 
\,.
\end{align}

\begin{result}
	\label{result:epsonetwo_sic}
	When the scheduler has selected a rate pair $(\rate_1,\rate_2)$ while assuming that user 2 is decoded first, the resulting decoding error probabilities are approximated as
\begin{align}
\epsone &\approx  Q\left(\frac{\snrturn-\snrmmse_1}{\sigmaicsione}\right)
+ \frac{\sigman}{2\sigmaicsione} e^{-\constc} \left( Q\left(\frac{\snrfactor_1-\mun}{\sigman} \right) - Q\left(\frac{\snrturn-\mun}{\sigman} \right) \right) +Q\left(\frac{\snrmmse_1-\snrfactor_1}{\sigmaicsione}\right)
\\\epstwo &\approx Q\left(\frac{\snrturn-\snrmmse_1}{\sigmaicsione}\right) 
+\frac{\sigman}{2\sigmaicsione} e^{-\constc} \left( Q\left(\frac{-\mun}{\sigman} \right) - Q\left(\frac{\snrturn-\mun}{\sigman} \right) \right)
\end{align}
with $\snrturn=\snrmmse_2/\snrfactor_2-1$ and
\begin{align}
\sigman^2 &= \sigmaicsione^2\left(1+\frac{\sigmaicsione^2 \snrfactor_2^2}{ \sigmaicsitwo^2}\right)^{-1}
\label{eq:sigman}
\\\mun &= \left(\snrmmse_1  +  \frac{\sigmaicsione^2}{\sigmaicsitwo^2}\snrfactor_2(\snrmmse_2-\snrfactor_2)   \right) 
\left(1+\frac{\sigmaicsione^2 \snrfactor_2^2}{\sigmaicsitwo^2}\right)^{-1}
\label{eq:mun}
\\\constc &= - \frac{1}{ 2\cdot \sigma_n^2}\mu_n^2 + \frac{1}{ 2\cdot \sigmaicsione^2}\left(\snrmmse_1^2 + \frac{\sigmaicsione^2}{ \sigmaicsitwo^2}\left( \snrmmse_2-\snrfactor_2\right)^2\right)
\label{eq:constn}
\end{align}
In the other case, when user 1 is decoded first, the resulting error probabilities can be obtained from the same expressions, after switching 1 and 2 in all expressions. 
\end{result}

\begin{proof}
The probability $\epsone$ that user 1 cannot be decoded is given as $\epsone = \epsonenointerf + \epstwodir$,
where $\epsonenointerf$ denotes the probability that signal 1 is in outage after applying SIC:
\begin{equation}
\epsonenointerf = \Prob{\Snr_1 < \snrfactor_1} \approx Q\left(\frac{\snrmmse_1-\snrfactor_1}{\sigmaicsione}\right)
\,,
\end{equation}
and $\epstwodir$ is the probability that signal 1 is not in outage $(\Snr_1 > \snrfactor_1)$, but signal 2 is in outage:
\begin{align}
\epstwodir
&=\int\limits_{\snr_1=\snrfactor_1}^{\infty} \Prob{ \frac{\Snr_2}{\snr_1+1} < \snrfactor_2} f_{\Snr_1}(\snr_1)d\snr_1
\label{eq:epstwodir_int1}
\\&\approx\int\limits_{\snr_1=\snrfactor_1}^{\infty} Q\left(\frac{\snrmmse_2-\snrfactor_2(\snr_1+1)}{\sigmaicsitwo}\right)  f_{\Snr_1}(\snr_1)d\snr_1
\end{align}
In order to obtain a closed-form approximation to this integral, we note that the argument of the Q-function is positive for $\snr_1<\snrturn$.
We then use the Chernoff bound $Q(x)\le\frac{1}{2}e^{-\frac{x^2}{2}}$ for $x\ge 0$ \cite{chiani2003exponential},
along with $Q(x) \le 1$ for $x<0$, to obtain
$\epstwodir  \approx \epstwodirlower+\epstwodirupper$
with
\begin{align}
\epstwodirlower&=\int\limits_{\snr_1=\snrfactor_1}^{\snrturn} \frac{1}{2}\exp\left(-\frac{(\snrmmse_2-\snrfactor_2(\snr_1+1))^2}{ 2 \cdot \sigmaicsitwo^2}\right)  
\frac{1}{\sqrt{2\pi\sigmaicsione^2}}\exp\left(-\frac{(\snr_1-\snrmmse_1)^2}{ 2\cdot \sigmaicsione^2}\right) d\snr_1 
\\&= \frac{1}{2} \frac{1}{\sqrt{2\pi\sigmaicsione^2}}\int\limits_{\snr_1=\snrfactor_1}^{\snrturn}\exp\left(-\frac{\left(\snr_1   - \mu_n  \right)^2}{ 2\cdot \sigma_n^2} - \constc\right)d\snr_1
\label{eq:eps_21a_magic_algebra}
\\&= \frac{\sigma_n}{2\sigmaicsione} e^{-\constc} \left( Q\left(\frac{\snrfactor_1-\mu_n}{\sigma_n} \right) - Q\left(\frac{\snrturn-\mu_n}{\sigma_n} \right) \right)
\,,
\end{align}
where \eqref{eq:eps_21a_magic_algebra} follows after tedious algebra, applying \eqref{eq:sigman}, \eqref{eq:mun}, and \eqref{eq:constn}.
We also find
\begin{align}
\epstwodirupper
=\int\limits_{\snrturn}^{\infty} f_{\Snr_1}(\snr_1)d\snr_1 \approx Q\left(\frac{\snrturn-\snrmmse_1}{\sigmaicsione}\right)
\,.
\end{align}

The error probability $\epstwo$ for user 2 is computed in the same way as $\epstwodir$, but the integral in \eqref{eq:epstwodir_int1} must start from zero.
\end{proof}

\begin{remark}
All approximation steps in the proof were designed towards creating upper bounds on the decoding error probabilities $\epsone$ and $\epstwo$. However, the derivations are based on the Gaussian approximation \eqref{eq:snr_gauss_approx} of the channel estimation errors, which may underestimate the interference. Therefore, the obtained analytical expressions are not strict upper bounds, but we validate numerically in Sec.~\ref{ssec:num_validate} that
the analytical expressions are usually either upper bounds or tight approximations for $\epsone$ and $\epstwo$.
\end{remark}

\subsection{Outage Probability under Joint Decoding}
\label{ssec:icsi_joint}
In the previous section, we considered a SIC receiver.
When using joint decoding, the receiver can decode both codewords when the rates are within the capacity region \eqref{eq:rate1}--\eqref{eq:sumrate}.
If any of the conditions are violated, a decoding error occurs. We assume that neither of the codewords can be decoded in that case.
The selected rates violate the capacity constraints with probabilities
\begin{align}
\varepsilon_\mathrm{I} &= \Prob{\rate_{1} > \log_2(1+\Snr_1)}
\\ \varepsilon_\mathrm{II} &= \Prob{\rate_{2} >\log_2(1+\Snr_2)}
\\ \varepsilon_\mathrm{III}&= \Prob{\rate_{1} + \rate_2 > \log_2(1+\Snr_1+\Snr_2)}
\end{align}
We follow \cite{schiessl2016imperfectcsi} and apply the Gaussian approximation \eqref{eq:snr_gauss_approx} to each term. We then obtain a closed-form bound on $\varepsilon$ by applying the union bound $\varepsilon \le \varepsilon_\mathrm{I}+ \varepsilon_\mathrm{II}+\varepsilon_\mathrm{III}$:
\begin{align}
\varepsilon &\le Q\left(\frac{\snrmmse_1-\snrfactor_1}{\sigmaicsione}\right)
+ 
Q\left(\frac{\snrmmse_2-\snrfactor_2}{\sigmaicsitwo}\right)
+ 
Q\left(\frac{\snrmmse_1+\snrmmse_2-2^{\rate_{1}+\rate_2}+1}{\sqrt{\sigmaicsione^2+\sigmaicsitwo^2}}\right)
\,.
\end{align}

%% file: finite_bl.tex
\section{Imperfect CSI and Finite-Length Coding}
\label{sec:icsi_fbl}
In the previous section, we used a simple outage model that assumes decoding errors occur if and only if the selected rates are above the channel capacity. However, when the blocklength $\nd$ of the channel code is finite, that model no longer holds. In fact, when the blocklength is finite, errors occur with probability $\varepsilon>0$ even when the channel is not in outage. In this section we analyze the joint impact of imperfect CSI and finite-length coding on the decoding error probabilities based on methods we developed in \cite{schiessl2016imperfectcsi}.

A well-known result for finite-length codes in AWGN channels with SNR $\snr$ is given by Polyanskiy et al. \cite{polyanskiy2010channel}, who showed in order to achieve an error probability $\varepsilon$, the transmitter should select a rate
\begin{equation}
\rate_\mathrm{AWGN}(\snr,\nd,\varepsilon) \approx \log_2(1+\snr) - \sqrt{\frac{\dispersionawgn(\snr)}{\nd}}Q^{-1}(\varepsilon)
\label{eq:rate_polyanskiy}
\,,
\end{equation}
where 
\begin{equation}
\dispersionawgn=\log_2^2(e)\left(1-\frac{1}{(1+\snr)^2}\right)
\label{eq:dispersion_awgn}
\end{equation}
is the channel dispersion. However, the codewords that achieve \eqref{eq:rate_polyanskiy} are not Gaussian distributed, which means that the result cannot be directly applied to the multiuser scenario, where non-Gaussian codewords would create non-Gaussian interference, which is hard to analyze.
When the codewords must be i.i.d. Gaussian, the achievable transmission rate is approximated as \cite{scarlett2017dispersion}
\begin{equation}
\rate_\mathrm{iid}(\snr,\nd,\varepsilon) \approx \log_2(1+\snr) - \sqrt{\frac{\dispersioniid(\snr)}{\nd}}Q^{-1}(\varepsilon)
\label{eq:rate_iid}
\,,
\end{equation}
which has the same form as \eqref{eq:rate_polyanskiy}, but with a different dispersion term
\begin{equation}
\dispersioniid=\log_2^2(e)\frac{2\snr}{1+\snr}
\label{eq:dispersion_iid}
\,.
\end{equation}
We now analyze the finite blocklength effects separately for the case of a decoder using successive interference cancellation, and for a joint decoder. The analysis in this section closely follows our previous works \cite{schiessl2016imperfectcsi,schiessl2019miso}. To simplify discussions, we assume that \eqref{eq:rate_iid} holds with equality.

\subsection{Error Probability for SIC Decoding}
Before analyzing the joint impact of imperfect CSI and finite blocklength coding, we first assume that the SNR values $\snr_1$ and $\snr_2$ are perfectly known. 
We again consider the case where the decoder uses successive interference cancellation and decodes user 2 first. In this case, the SINR of the signal is given as $\sinr_2=\frac{\snr_2}{\snr_1+1}$. Due to the assumption of i.i.d. Gaussian codewords for user 1, the signal is equivalent to that of an AWGN channel with SNR $\sinr_2$. Thus, the achievable rate for user 2 can be determined by \eqref{eq:rate_iid}. We can solve \eqref{eq:rate_iid} for $\varepsilon$
to obtain the decoding error probability given the selected rate $\rate_2$:
\begin{equation}
\varepsilon_2 = Q\left(\frac{\log_2\left(1+\sinr_2\right) - \rate_2}{\dispersioniid\left(\sinr_2\right)/\nd}\right)
\label{eq:eps_two_fbl}
\,.
\end{equation}
Following our previous work \cite{schiessl2016imperfectcsi}, we define the random \emph{blocklength-equivalent capacity}
\begin{equation}
C_\fblequivtwo =\log_2\left(1+\sinr_2\right) + \sqrt{\frac{\dispersioniid\left(\sinr_2\right)}{\nd}}\Ubtwo
\label{eq:cap2_fblequiv}
\,
\end{equation}
with $\Ubtwo\sim\mathcal{N}(0,1)$ and observe for a given $\sinr_2$ that $\varepsilon_2$ is (by definition) the outage probability of a channel with random capacity $C_\fblequivtwo$: $\varepsilon_2=\Prob{C_\fblequivtwo < \rate_2}$.
Similarly, we define $\Ubone\sim\mathcal{N}(0,1)$ and find that the error probability $\epsonenointerf$ of user 1, after SIC was applied, is given as $\epsonenointerf=\Prob{C_\fblequivone < \rate_1}$ with
\begin{equation}
C_\fblequivone =\log_2\left(1+\snr_1\right) + \sqrt{\frac{\dispersioniid\left(\snr_1\right)}{\nd}}\Ubone
\label{eq:cap1_fblequiv}
\,.
\end{equation}

All of the above statements remain true when the instantaneous SNR $\snr_k$ is not perfectly known at the transmitters, i.e., we can replace $\snr_k$ with $\Snr_k$, where all random variables $U_k$ and $\Snr_k$ are mutually independent. 
At the receiver, we still assume perfect CSI, as motivated in Sec.~\ref{sec:icsi}.
For user 1, we can directly follow our previous work \cite{schiessl2016imperfectcsi} to obtain a bound on $\epsonenointerf$. Using a first-order Taylor approximation, we obtain the bound
\begin{align}
\epsonenointerf&\le\Prob{\log_2\left(1+\Snr_1 + \sigmafblone(\Snr_1) \Ubone \right) < \rate_1}
\end{align}
with 
\begin{align}
\sigmafblone(\Snr_1) =\frac{1+\Snr_1}{\log_2(e)} \sqrt{\frac{\dispersioniid\left(\Snr_1\right)}{\nd}}
\label{eq:icsi_fbl_sigmafblone}
\,.
\end{align}
Using the Gaussian approximation \eqref{eq:snr_gauss_approx} for $\Snr_1$, and replacing $\sigmafblone(\Snr_1)$ with its estimated value $\sigmafblone(\snrmmse_1)$, we obtain \cite{schiessl2016imperfectcsi}
\begin{align}
\epsonenointerf \approx Q\left(\frac{\snrmmse_1-\snrfactor_1}{\sigmaicsifblone(\snrmmse_1)}\right)
\label{eq:icsi_fbl_approx_tcom}
\,,
\end{align}
with the variance of the sum of the independent Gaussian variables $\Snr_1$ and $\Ubone$ given as
\begin{align}
\sigmaicsifblone(\snrmmse_1) = \sqrt{\sigmaicsione^2+\sigmafblone^2(\snrmmse_1)}
\,,
\end{align}
where $\sigmaicsione^2$ is given by \eqref{eq:sigmaicsi}.
User 2 is decoded directly, with error probability
\begin{align}
\varepsilon_2&=\Prob{C_\fblequivtwo < \rate_2}
\\&=\Prob{\log_2\left(1+\frac{\Snr_2}{\Snr_1+1}\right) + \sqrt{\frac{\dispersioniid\left(\frac{\Snr_2}{\Snr_1+1}\right)}{\nd}}\Ubtwo< \rate_2}
\\&\le\Prob{\log_2\left(1+\frac{\Snr_2}{\Snr_1+1} + \sigmafbltwo\left(\frac{\Snr_2}{\Snr_1+1}\right) \Ubtwo\right) < \rate_2}
\label{eq:eps2_deriv_taylor}
\\&=\Prob{\Snr_2 + (1+\Snr_1)\sigmafbltwo\left(\frac{\Snr_2}{\Snr_1+1}\right) \Ubtwo< \snrfactor_2(1+\Snr_1)}
\\&\approx\Prob{\Snr_2 + (1+\snrmmse_1)\sigmafbltwo\left(\frac{\snrmmse_2}{\snrmmse_1+1}\right) \Ubtwo< \snrfactor_2(1+\Snr_1)}
\label{eq:eps2_deriv_expectedapprox}
\,,
\end{align}
where \eqref{eq:eps2_deriv_taylor} follows again from a Taylor approximation, and in \eqref{eq:eps2_deriv_expectedapprox} we replaced the random variables in and before $\sigmafbltwo$ with their estimated values, similar to \cite{schiessl2016imperfectcsi}.
Thus:
\begin{align}
\epstwo&\approx\int\limits_{\snr_1=0}^{\infty} \Prob{ \frac{\Gicsifbltwo}{\snr_1+1} < \snrfactor_2} f_{\Snr_1}(\snr_1)d\snr_1
\label{eq:epstwodir_int1_new}
\end{align}
where $\Gicsifbltwo=\Snr_2 + (1+\snrmmse_1)\sigmafbltwo\left(\frac{\snrmmse_2}{\snrmmse_1+1}\right) \Ubtwo$ describes the uncertainty of signal 2 due to imperfect CSI and finite blocklength. $\Gicsifbltwo$ is Gaussian with variance
\begin{align}
\sigmanewicsifbltwo^2(\snrmmse_1,\snrmmse_2)=\sigmaicsitwo +  (1+\snrmmse_1)^2\sigmafbltwo^2\left(\frac{\snrmmse_2}{\snrmmse_1+1}\right)
\end{align}
Following the same steps as in Sec.~\ref{ssec:icsi_cornerpoints}, we obtain
\begin{align}
\epstwo &\approx Q\left(\frac{\snrturn-\snrmmse_1}{\sigma_1}\right) 
+\frac{\sigman}{2\sigma_1} e^{-\constc} \left( Q\left(\frac{-\mun}{\sigman} \right) - Q\left(\frac{\snrturn-\mun}{\sigman} \right) \right)
\end{align}
with $\sigman$, $\mun$, and $\constc$ still given by \eqref{eq:sigman}, \eqref{eq:mun}, and \eqref{eq:constn}, but with $\sigma_{2}$ replaced by $\sigmanewicsifbltwo$.

Finally, user 1 cannot be decoded when it cannot be decoded after interference cancellation or when user 2 cannot be decoded. Contrary to the analysis with infinite blocklength in Sec.~\ref{sec:icsi}, those events may not be independent,
so we have to apply the union bound:
\begin{align}
\epsone \le \epsonenointerf + \epstwo
\,.
\end{align}

\subsection{Error Probability for Joint Decoding}
The achievable rate region for a multiple access channel with finite blocklength coding has been studied by MolavianJazi \cite{molavianjazi2014phdthesis}.
We first assume that the SNR values $\snr_1$, $\snr_2$ are perfectly known, such that these values correspond to the power constraints on the codewords.
It was shown in \cite[Thm.~7]{molavianjazi2014phdthesis} that given a maximum error probability $\varepsilon$, a second-order approximation for the achievable rate region can be found by splitting the error probability into three arbitrarily large parts with $\varepsilon = \varepsilon_\mathrm{I} + \varepsilon_\mathrm{II} + \varepsilon_\mathrm{III}$. The rates $\rate_1,\rate_2$ are achievable with error probability $\varepsilon$ if
\begin{align}
\rate_{1} &\le \log_2(1+\snr_1) - \sqrt{\dispersionawgn(\snr_1)/\nd}\cdot Q^{-1}(\varepsilon_\mathrm{I}) + \mathcal{O}\left(1/\nd\right)
\label{eq:fbl_joint_achrate1}
\\ \rate_{2} &\le \log_2(1+\snr_2)- \sqrt{\dispersionawgn(\snr_2)/\nd} \cdot Q^{-1}(\varepsilon_\mathrm{II}) + \mathcal{O}\left(1/\nd\right)
\label{eq:fbl_joint_achrate2}
\\ \rate_{1} + \rate_2 &\le \log_2(1+\snr_1+\snr_2)- \sqrt{\dispersionmac(\snr_1,\snr_2)/\nd}\cdot Q^{-1}(\varepsilon_\mathrm{III}) + \mathcal{O}\left(1/\nd\right)
\label{eq:fbl_joint_achrate3}
\end{align}
with $\dispersionawgn$ given in \eqref{eq:dispersion_awgn} and
\begin{align}
\dispersionmac(\snr_1,\snr_2) = \dispersionawgn(\snr_1+\snr_2) + 2\log_2^2(e)\frac{\snr_1\snr_2}{(1+\snr_1+\snr_2)^2}
\,.
\end{align}
Like in the previous section, we assume that the second-order approximations are exact, i.e., we ignore the terms $\mathcal{O}\left(1/\nd\right)$. Then, we define random blocklength-equivalent capacities
\begin{align}
C_\fblequivonenew &= \log_2(1+\snr_1) + \sqrt{\dispersionawgn(\snr_1)/\nd}\cdot U_\mathrm{I} 
\\ C_\fblequivtwonew &= \log_2(1+\snr_2)+ \sqrt{\dispersionawgn(\snr_2)/\nd}\cdot U_\mathrm{II}
\\ C_\fblequivcomb &= \log_2(1+\snr_1+\snr_2) +  \sqrt{\dispersionmac(\snr_1,\snr_2)/\nd}\cdot U_\mathrm{III}
\end{align}
with $U_\mathrm{I},U_\mathrm{II},U_\mathrm{III}\sim\mathcal{N}(0,1)$. 
If we redefine $\varepsilon_\mathrm{I}$, $\varepsilon_\mathrm{II}$, and $\varepsilon_\mathrm{III}$ as the probabilities that the rates exceed $C_\fblequivonenew$, $C_\fblequivtwonew$, and $C_\fblequivcomb$, respectively, then the overall probability that the rates $(\rate_1,\rate_2)$ are outside the random blocklength-equivalent capacity region is bounded by $\varepsilon = \varepsilon_\mathrm{I} + \varepsilon_\mathrm{II} + \varepsilon_\mathrm{III}$. In other words, our definitions lead to the same relationship between $(\rate_1,\rate_2)$ and $\varepsilon$ as in \cite[Thm.~7]{molavianjazi2014phdthesis}.
The codewords that achieve \eqref{eq:fbl_joint_achrate1} to \eqref{eq:fbl_joint_achrate3} are chosen independently of each other, and are uniformly distributed on the ``power shells" \cite{molavianjazi2014phdthesis}.
Therefore, the choice of codewords depends only on the selected rates, but not on the fading state, and we can thus extend the above results
directly to fading channels. 
For $\varepsilon_\mathrm{I}$ and $\varepsilon_\mathrm{II}$, we can then directly use the approximation \eqref{eq:icsi_fbl_approx_tcom}, but using a dispersion term $\dispersionawgn$.
Using similar steps, we obtain
$
\varepsilon_\mathrm{III}\approx Q\left(\frac{\snrmmse_1+\snrmmse_2-\snrfactor_{12}}{\sigma_\mathrm{III}}\right)
$
with
\begin{align}
\sigma_\mathrm{III} = \sqrt{\sigmaicsione^2+\sigmaicsitwo^2+\frac{(1+\snrmmse_1+\snrmmse_2)^2}{\log_2^2(e)}\frac{\dispersionmac(\snrmmse_1,\snrmmse_2)}{\nd}}
\,.
\end{align}

%% file: Numerics.tex

\section{Numerical Results}
\label{sec:numerics}
In this section, we evaluate the delay performance of NOMA and OMA schemes. Furthermore, we quantify the impact of imperfect CSI and finite-length coding on the delay performance.
First, we present in Sec.~\ref{ssec:numerics:general} the general methodology used in the evaluation. 
The performance of NOMA depends on the rate adaptation, and we thus show in Sec.~\ref{ssec:numerics_ideal} how different rate adaptation schemes chosen by the base station result in different delay violation probabilities for the two users, assuming an ideal system model. 
In Sec.~\ref{ssec:num_validate}, we consider the effects of imperfect CSI and validate the outage probability approximations from Sec.~\ref{ssec:icsi_cornerpoints}. In Sec.~\ref{ssec:numerics_icsi}, we analyze the queueing performance of the system under different delay constraints, where we also evaluate the performance impact due to imperfect CSI. Finally, Sec.~\ref{ssec:numerics_fbl} shows results when both imperfect CSI and finite-length coding are taken into account.

\subsection{Metrics and Methodology}
\label{ssec:numerics:general}
For comparing different schemes, we generally use the upper bound \eqref{eq:pdelay_bound} on the delay violation probability $\pvk(w)$. The delay performance depends on the number of bits $\alpha_k$ that arrive in each user's queue per time slot. To compare the delay performance of different schemes over a wide range of parameters, we will often assume fixed target delay parameters (e.g., $\pvk(5)<10^{-8}$) and show the maximum arrival rates $\alpha_1$, $\alpha_2$ such that these constraints can still be met.

\subsubsection{Optimization Method}
We quantize the distribution of the estimated SNR values $\snrmmse_1,\snrmmse_2$ until finer quantization yields no more significant improvement, usually around 300 points each. For the ideal model with perfect CSI ($\snr_k=\snrmmse_k$), we can determine the optimal rates according to the results in Sec.~\ref{sec:analysis}.
For imperfect CSI and finite blocklength, we consider several rate pairs $(\rate_{1,i},\rate_{2,i})$ for each combination of quantized $(\snrmmse_{1,i},\snrmmse_{2,i})$, each resulting in different approximate error probabilities $\varepsilon_{1,i}$ and $\varepsilon_{2,i}$. Then, for fixed values $\s_1$, $\s_2$, and $\tilde{\lambda}$, we can determine the optimal rates according to \eqref{eq:argmin_rates_discrete}.
Instead of an exhaustive search over all combinations of $\s_1$, $\s_2$, and $\tilde{\lambda}$, we propose a suboptimal approach where we start with a coarsely quantized grid for $\s_k$ and $\tilde{\lambda}$.
We then iterate between optimizing the rates and choosing $\s_k$ and $\tilde{\lambda}$ such that the kernels $\mathcal{K}_k\left(\s_K,w\right)$ are minimized. This method seems to offer robust and reasonably fast convergence.

\subsubsection{Comparison to OMA and Power Allocation}
We assume throughout this section that the signals are subject to a constraint on the sum instantaneous transmit power. We denote the average SNR when one device uses the entire transmit power as $\snravgoma_1$ and $\snravgoma_2$, respectively.
These are also the average SNR values of the signals when using OMA.
For NOMA, both devices are active, so the transmit powers must be scaled by factors $\powfac_1$ and $\powfac_2$, with $\powfac_1+\powfac_2=1$, so that the average SNR values are given as $\snravg_1=\powfac_1\snravgoma_2$ and $\snravg_2=\powfac_2\snravgoma_2$. We always consider scenarios where $\snravgoma_1>\snravgoma_2$, and assign $\powfac_1=0.2$ and $\powfac_2=0.8$.
During the channel training phase, channel access is orthogonal, so the SNR during training is $\snravgoma_1$ and $\snravgoma_2$, respectively.

\subsection{Results for Ideal Model}
\label{ssec:numerics_ideal}
\begin{figure}[t!]
	\centering
	\subfloat[\label{subfig:pcsi_optrates}]{%
		\includegraphics[width=0.93\figurewidth]{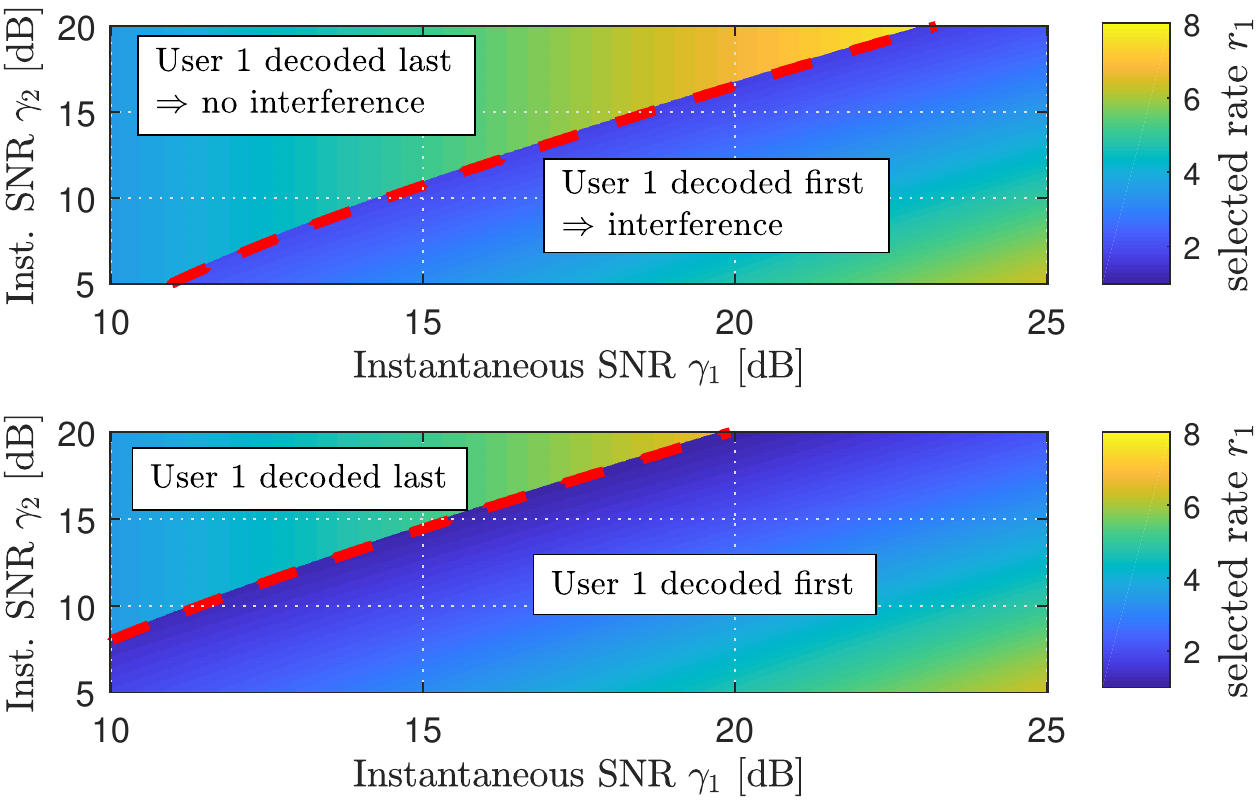}
	}
	\subfloat[\label{subfig:pcsi_delayviol}]{%
		\includegraphics[width=0.93\figurewidth]{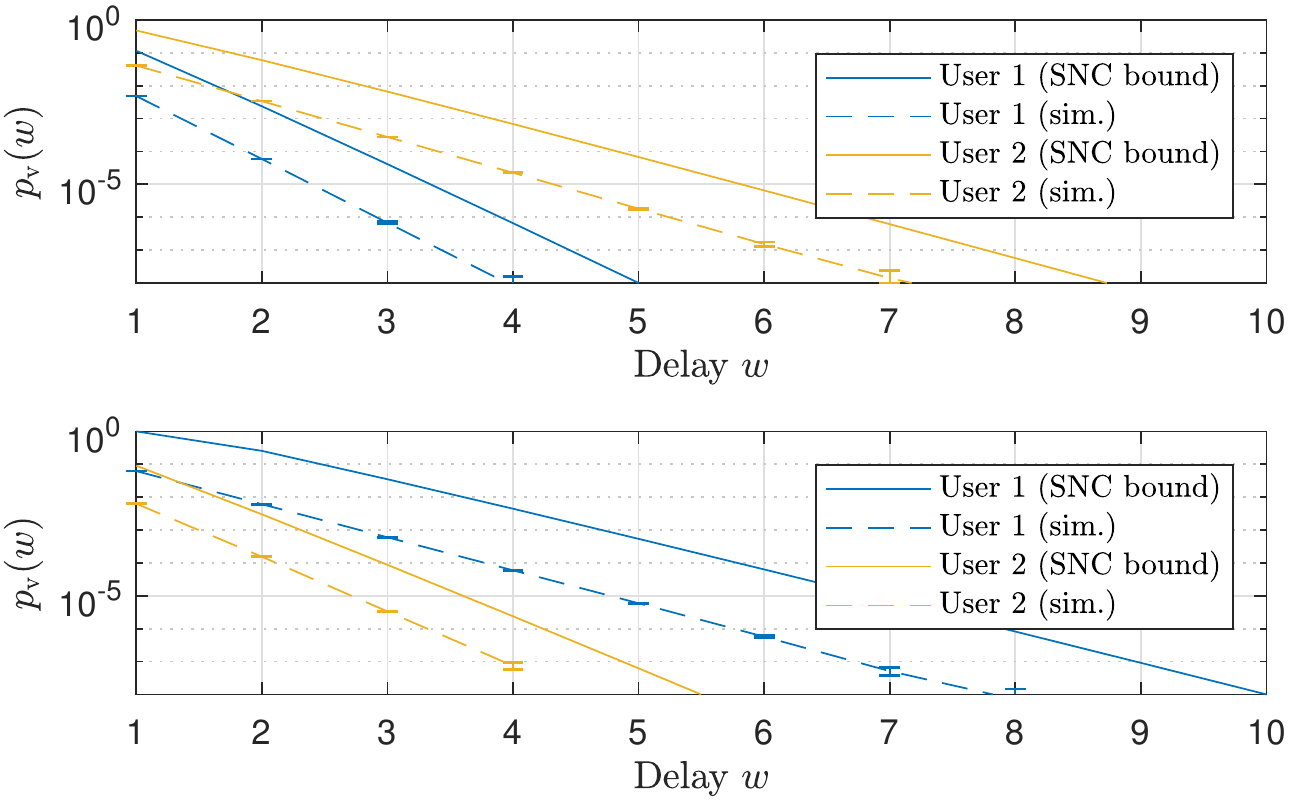}
	}
	\vfill
	\caption{Average SNR: $\snravgoma_1=30~\mathrm{dB}$, $\snravgoma_2=15~\mathrm{dB}$. $\ntotal=250$, $\nd=200$, $\alpha_1=560$, $\alpha_2=320$ bits per time slot. Two different rate adaptations $\rateadaptvec$ (top/bottom).  a) Optimal rate $r_1=\rateadapt_1(\snr_{1},\snr_2)$. b) Delay violation probability $\pv(w)$ (From SNC bounds and from simulations, 95\% confidence intervals shown). }
	\label{fig:results_pcsi_rateadapt}
\end{figure}

First, we consider SIC decoding and assume an ideal system model where the CSI at the transmitter is perfect and finite blocklength effects are ignored. 
In Fig.~\ref{fig:results_pcsi_rateadapt}, we study the effect of different rate adaptation functions $(\rate_1,\rate_2)=\rateadaptvec(\snr_1,\snr_2)$ on the delay performance of both users under SIC. In the upper plots, the function $\rateadaptvec$ is chosen such that the delay violation probability $\pv(w)$ of the second user is minimized, subject to a constraint $\pv(w=5)<10^{-8}$ for the first user. The top of Fig.~\ref{subfig:pcsi_optrates} shows the optimal $\rate_1=\rateadapt_1(\snr_1,\snr_2)$ under this constraint, while the top of Fig.~\ref{subfig:pcsi_delayviol} shows the resulting delay violation probabilities $\pv(w)$ for both users. The stochastic network calculus (SNC) bound on $\pv(w)$ for the first user exactly matches the constraint of $10^{-8}$ for a delay of $w=5$. The delay performance of the second user is worse. The bottom of Fig.~\ref{subfig:pcsi_optrates} shows the optimal $\rate_1=\rateadapt_1(\snr_1,\snr_2)$ under a different delay constraint with target delay $w=10$ for the first user, while the bottom of Fig.~\ref{subfig:pcsi_delayviol} shows $\pv(w)$ for that rate adaptation scheme. 
Fig.~\ref{subfig:pcsi_delayviol} shows that by increasing the maximum tolerable delay for the first user, we have reduced the delay violation probabilities for the second user. The heat maps in Fig.~\ref{subfig:pcsi_optrates} show that in the upper-left region, where the signal from user 1 is weak but the interference is strong, user 1 is decoded last, i.e., without interference. The size of this region shrinks when we improve the delay performance of user 2 at the expense of user 1.

We also determined the actual delay violation probabilities $\pv(w)$ empirically through Monte-Carlo simulations of the queueing system over $10^{10}$ time slots. As expected, the simulated $\pv(w)$ is below the analytical upper bounds from SNC. There is a gap of one or two orders of magnitude between the simulations and the upper bounds, which was observed also in previous works on SNC \cite{alzubaidy2016ton,schiessl2015delay,schiessl2016imperfectcsi,petreska2015recursive}. 
Regardless of the gap, the slopes of the simulation curves match the slopes of the analytical curves, and the bounds accurately predict how different rate adaptations will affect the delay performance of both users. 
Thus, the proposed analysis using SNC is sufficiently accurate to determine the optimal trade-offs between the two users.


\subsection{Validating the Approximations for Imperfect CSI}
\label{ssec:num_validate}
\begin{figure}[t!]
	\centering
	\subfloat[\label{subfig:verify_eps12_sic}]{
		\includegraphics[width=0.93\figurewidth]{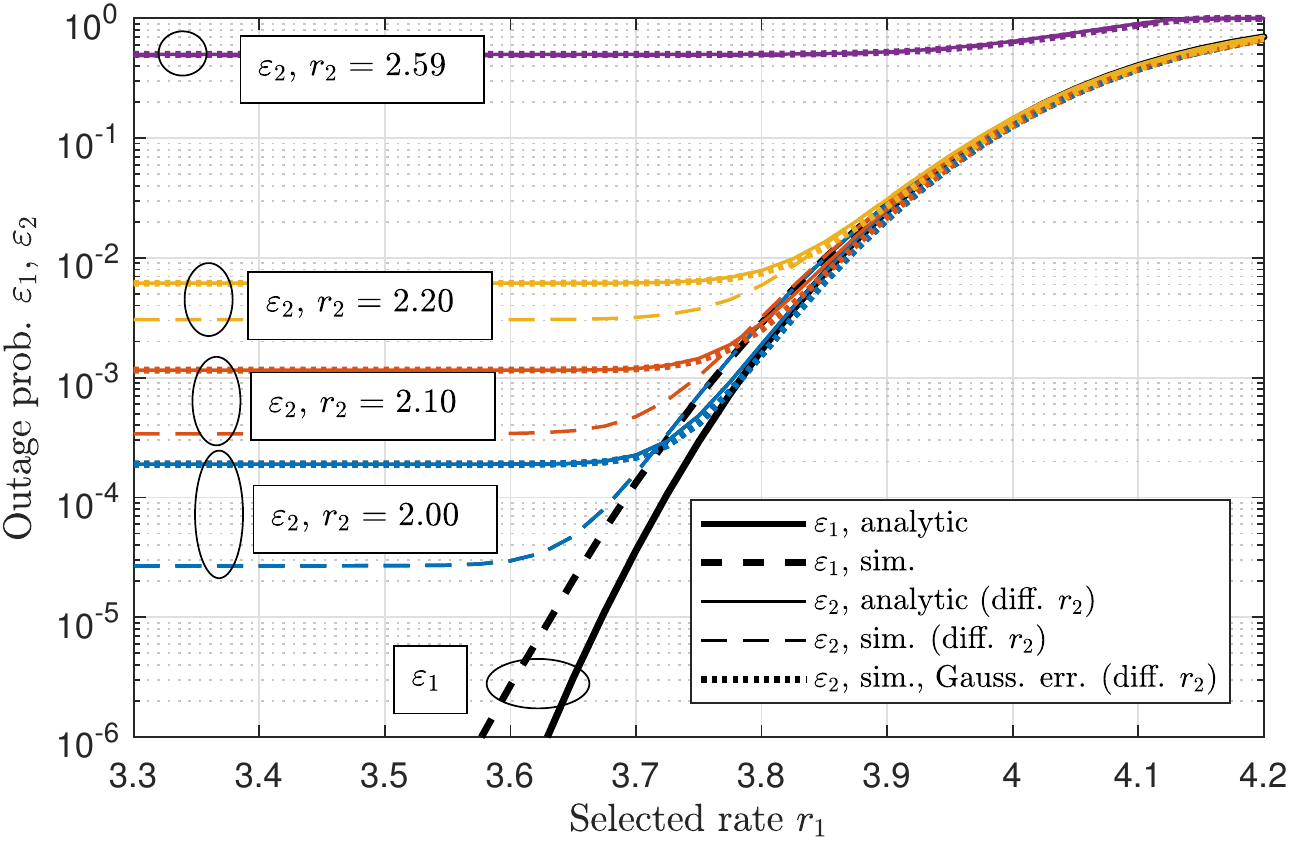}
	}
	\subfloat[\label{subfig:verify_sic_queueing}]{%
		\includegraphics[width=0.93\figurewidth]{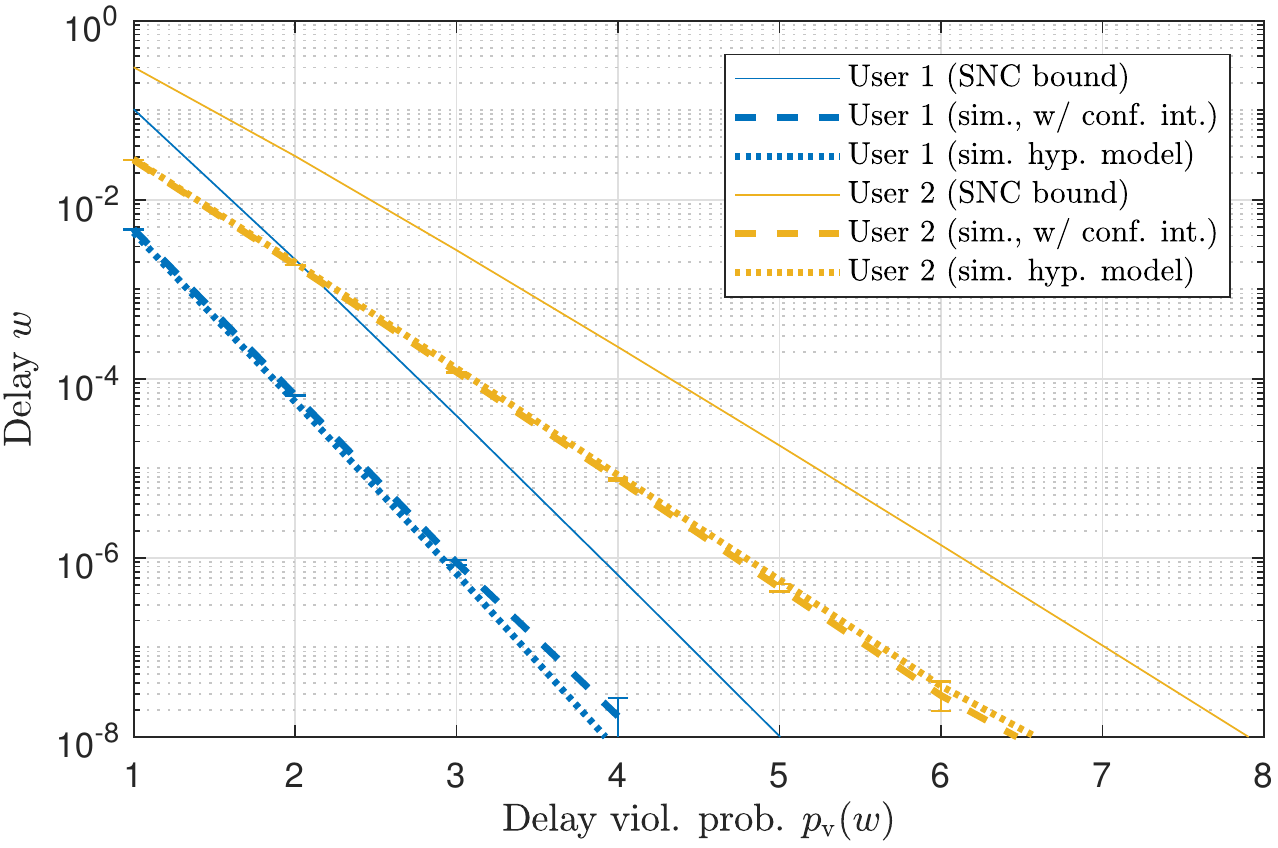}
	}
	\caption{Average SNR: $\snravgoma_1=30~\mathrm{dB}$,  $\snravgoma_2=15~\mathrm{dB}$. $\ntotal=250$ $\ntk{1}=\ntk{2}=25$. a) Estimated SNR: $\snrmmse_1=20~\mathrm{dB}$, $\snrmmse_2=7~\mathrm{dB}$. Decoding error probabilities $\epsone$, $\epstwo$ according to analytical results and simulations, vs. $\rate_1$ for different $\rate_2$. b) $\snravg_1=20~\mathrm{dB}$, $\snravg_2=10~\mathrm{dB}$, Delay viol. prob. $\pv(w)$ for a system  where the analytical $\epsone$, $\epstwo$ were used for optimal rate adaptation.}
	\label{fig:verify_sic}
\end{figure}

When considering that the rate adaptation mechanism only has access to an imperfect estimate of the channel state, outages may occur. For such a scenario, we perform the delay analysis and rate optimization for NOMA with SIC decoding based on the analytical approximations for the outage probabilities $\epsone$ and $\epstwo$ that were derived in Sec.~\ref{ssec:icsi_cornerpoints}. First, we verify that these analytical approximations are sufficiently accurate for the rate adaptation $(\rate_1,\rate_2)=\rateadaptvec(\snrmmse_1,\snrmmse_2)$.
In Fig.~\ref{subfig:verify_eps12_sic}, we plot both $\epsone$ and $\epstwo$ when the base station estimates the instantaneous SNRs as $\snrmmse_1=20~\mathrm{dB}$ and $\snrmmse_2=7\mathrm{dB}$ and tries to adapt the coding rates $(\rate_1,\rate_2)$ according to these estimates. User 1 is decoded first. When the message from user 1 cannot be decoded (which depends on $\rate_1$), the signal cannot be removed, and the SIC decoding fails. Otherwise, the base station will try to decode signal 2, which can again fail, depending on $\rate_2$. We note that the analytical approximations for $\epsone$ (solid curves) are close to the actual error probabilities (dashed curves), obtained from Monte Carlo simulations with $10^8$ trials. This is important for rate adaptation: if the rate adaptation scheme would ignore the imperfections in the channel estimates, and select a rate of $\rate_1=\log_2(1+\snrmmse_1/(1+\snrmmse_2))\approx 4.14$ based on the estimates, then the system would experience error probabilities $\epsone$ of around 50\%. If the rate selection wants to keep $\epsone$ below $10^{-3}$, one would actually need to choose $\rate_1\approx 3.76$ according to the Monte-Carlo simulations. When using the analytical approximation, one would select approximately the same rate ($\rate_1\approx 3.78$). Thus, one can perform accurate rate adaptation without the need for extensive simulations (or numerical integrations). For $\epstwo$, we notice a fairly large gap between the analytical approximations of $\epstwo$ and the actual $\epstwo$ at $\rate_2=2.0$. This gap is caused by the Gaussian approximation for the channel estimation error. When we assume a hypothetical model where the last term in \eqref{eq:snr_snrmmse_equation} is ignored, i.e., where the channel estimation error is exactly Gaussian (dotted curves), then the gap between the approximation and the actual $\epstwo$ becomes zero. Nevertheless, the gap due to the Gaussian approximation is not harmful for the rate adaptation. When considering systems with a deadline of e.g. $w=5$ time slots, then individual decoding error probabilities in the range  $10^{-3}$ to $10^{-2}$ are sufficient to transmit the packets within the deadline. In this range, the Gaussian approximation is sufficiently accurate for the delay analysis, which we confirmed for the single-user scenario in \cite{schiessl2016imperfectcsi}. 
Most importantly, the approximations are either very close to the actual $\epsone$ or they overestimate $\epsone$. When the base station selects the rates based on an overestimation of the error probability, it will make a choice that is too conservative and select a slightly smaller rate than necessary, leading to a smaller error probability.

In Fig.~\ref{subfig:verify_sic_queueing}, we verify the accuracy of the approximations based on the resulting delay violation probabilities $\pv(w)$. The rate adaptation scheme was optimized using the analytical approximations for $\epsone$ and $\epstwo$, and the resulting analytical bounds on $\pv(w)$ are shown. The dashed curves show the results from simulating the queueing system according to the actual system model, while the dotted curves show $\pv(w)$ for a hypothetical system which exactly matches the analytical approximations (i.e., in the simulation, we generate outage events as Bernoulli random variables with the same probabilities as in the analysis). There is a gap between both simulated curves and the analytical results, but such a gap was already observed in case of perfect CSI in Fig.~\ref{subfig:pcsi_delayviol}, i.e., the gap is most likely due to the conservativeness of the SNC bounds. On the other hand, there is no significant difference between the actual system model and the hypothetical/approximate model, which shows that the actual error probabilities $\epsone$ and $\epstwo$ and their analytical approximations are very close for the parameters considered in the queueing analysis. In other words, whether we consider the actual system model or our analytical approximations, the queueing performance remains the same. We conducted further experiments which show that the analytical approximations work very well when the average SNR $\snravg_{k}$ for each user's signal is above $10~\mathrm{dB}$ and for training sequence length $\ntk{k}$ is above $25$. 

\hlchangeb{In the above Fig.~\ref{fig:verify_sic}, we have shown the evaluations for the approximate outage probabilities under SIC decoding and assuming infinite blocklength. The approximations for joint decoding are also based on the Gaussian approximation for the estimation error, which we have now shown to be accurate enough for the relevant parameters. Lastly, we performed additional simulations to validate the approximations for finite-length coding and found that $\pv(w)$ for the actual system model was slightly below the $\pv(w)$ for the hypothetical/approximate model. This means that although the approximations can be slightly inaccurate, they act as upper bounds on the error probability, similar to the approximation in \cite{schiessl2016imperfectcsi}. The actual system performance is then even better than predicted.
}

\subsection{Effects of Imperfect CSI}
\label{ssec:numerics_icsi}
\begin{figure}[t!]
	\centering
	\subfloat[\label{subfig:asymm_pcsi}]{%
		\includegraphics[width=0.95\figurewidth]{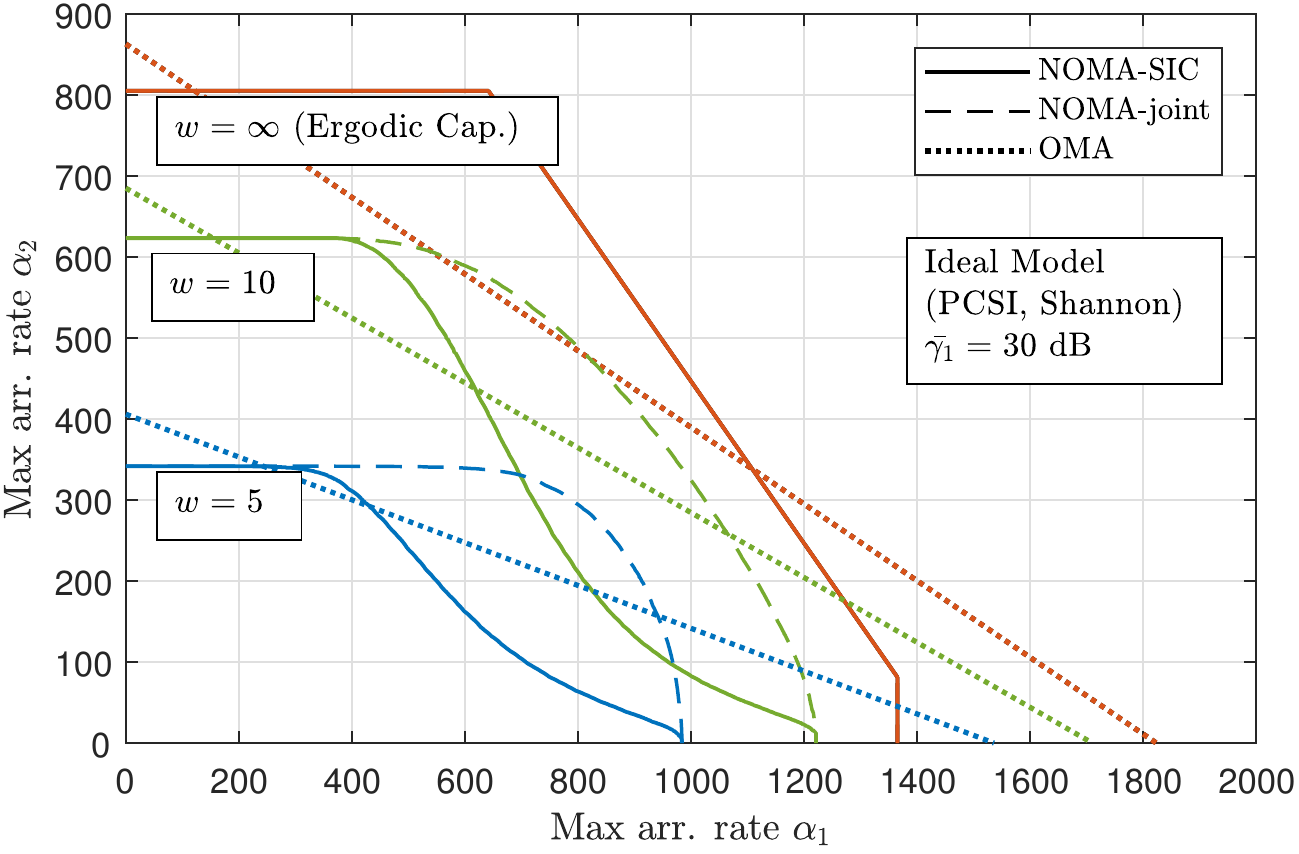}
	}
	\subfloat[\label{subfig:asymm_icsi}]{%
		\includegraphics[width=0.95\figurewidth]{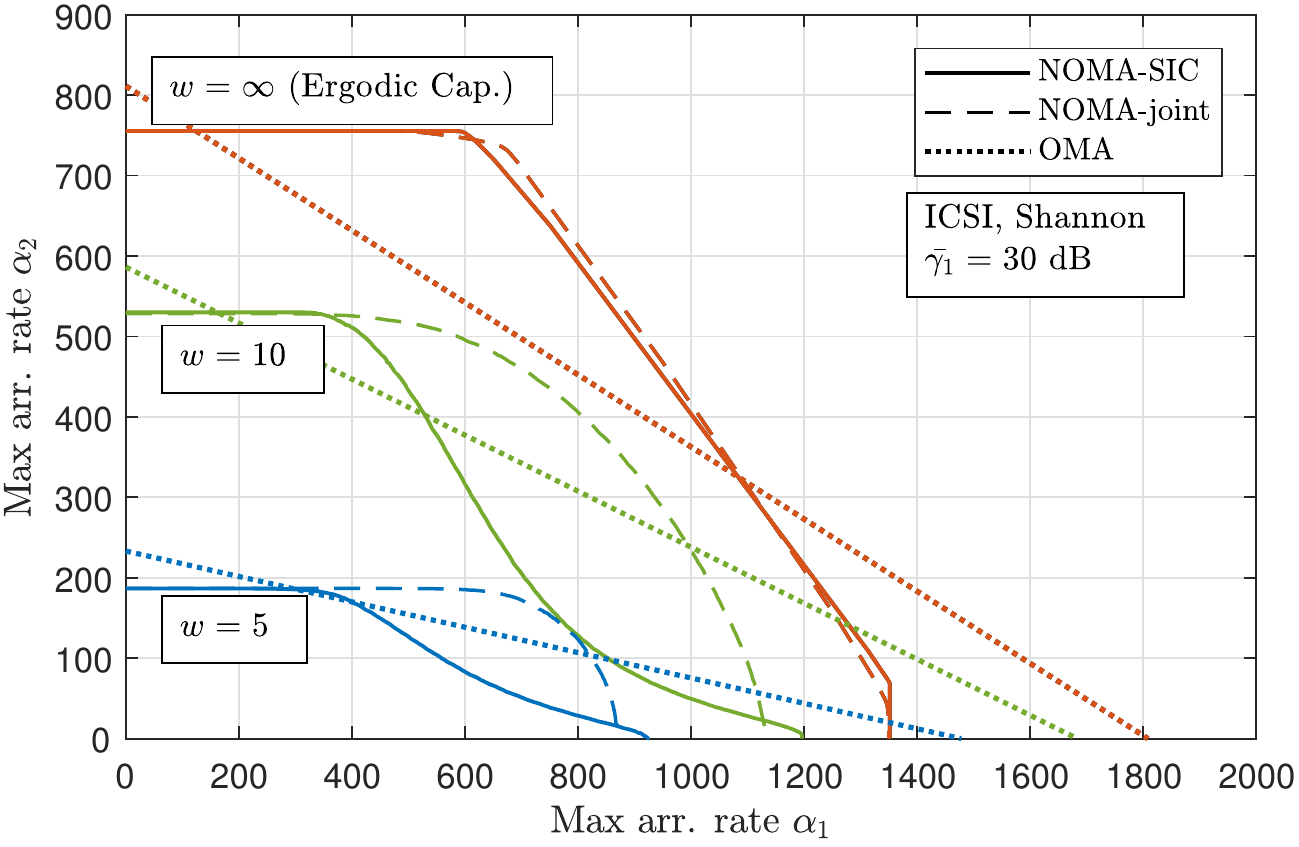}
	}
	\vfill
	\caption{Max arrival rates $\alpha_1$, $\alpha_2$ in bits per time slot s.t. $\pv(w)<10^{-8}$ for different $w$. Average SNR: $\snravgoma_1=30~\mathrm{dB}$, $\snravgoma_2=15~\mathrm{dB}$. $\ntotal=250$, $\ntk{1}=\ntk{2}=25$, $\nd=200$. a) Perfect CSI b) Imperfect CSI.}
	\label{fig:results_asymm}
\end{figure}

In Fig.~\ref{fig:results_asymm}, we investigate the performance for the ideal model with perfect CSI, and compare the resulting performance also to a semi-realistic model with imperfect CSI. In both cases, we still assume that errors occur only when the rates are above the Shannon capacity (i.e., infinitely long codewords). 
First, Fig.~\ref{subfig:asymm_pcsi} shows results for the ideal model with perfect CSI. 
The uppermost curves show the ergodic capacity per time slot, which corresponds to the maximum supported arrival rates $\alpha_1$ and $\alpha_2$ when there is no delay constraint (the delay may be infinite).
Then, we investigate the maximum arrival rates $\alpha_1$ and $\alpha_2$ such that, with an optimized rate adaptation scheme, the system can still meet delay constraints $\pv(w)<10^{-8}$ for both users, for different target delays $w$. 
We find that imposing tight delay requirements ($w=5$) degrades the performance, the maximum arrival rate reduces drastically compared to the ergodic case. 
For a maximum delay $w=5$, NOMA with SIC decoding now performs significantly worse than NOMA with joint decoding, because the optimal rate points for joint decoding often lie between the corner points of the capacity region. With SIC decoding, the rate adaptation scheme can only select the suboptimal corner points. Interestingly, for NOMA with joint decoding, the performance of the second user remains constant over a wide range of arrival rates $\alpha_1$ for the first user, i.e., both users can simultaneously achieve a large fraction of the maximum performance. 
\hlchange{For both system models, NOMA-joint significantly outperforms the orthogonal scheme (OMA), except for the regions where either $\alpha_1$ or $\alpha_2$ are very small. We found through further experiments that the performance of NOMA in those regions could be improved by using different power allocations $\powfac_1$,$\powfac_2$, from which we conclude that NOMA-joint always outperforms OMA for the considered scenario with $w=5$. }
On the other hand, NOMA with SIC decoding can only provide a small performance improvement over OMA for $w=5$. 
\hlchange{For different power allocations, we found that NOMA-SIC may not even achieve same performance as OMA.}

In Fig.~\ref{subfig:asymm_icsi}, we show results for the same parameters as in Fig.~\ref{subfig:asymm_pcsi}, but considering imperfect CSI.
We note first of all that imperfect CSI barely affects the performance under loose delay constraints ($w=\infty$ and $w=10$), but reduces the maximum achievable $\alpha_2$ under tighter delay constraints ($w=5$) by more than 40\%.
For both models, we observe that NOMA-joint significantly outperforms OMA. However, when considering imperfect CSI, NOMA-SIC can no longer outperform OMA for the considered parameters. We conclude that imperfect CSI creates a slightly larger performance penalty for NOMA-SIC than for OMA.

\begin{figure}[ht!]
	\centering
	%
	\subfloat[\label{subfig:asymm_symm_icsi}]{%
		\includegraphics[width=0.95\figurewidth]{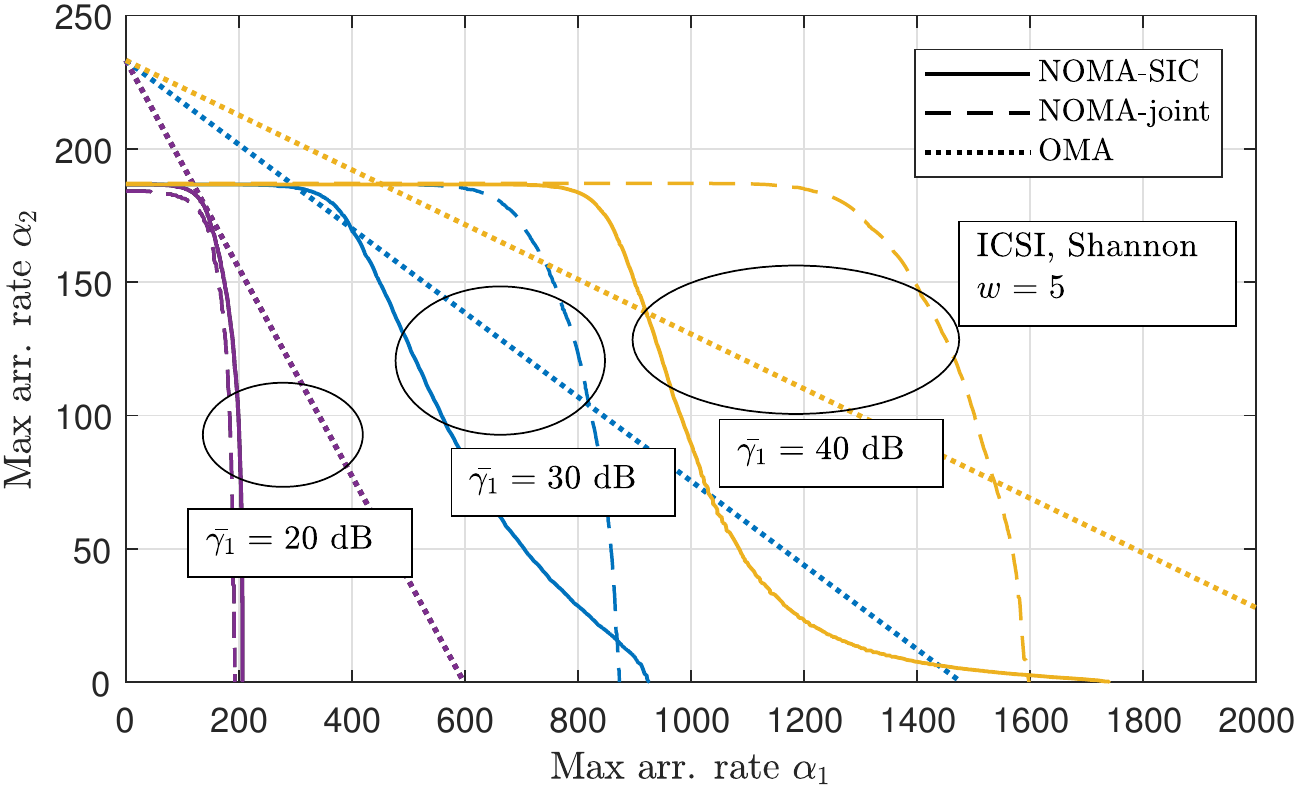}
	}
	\subfloat[\label{subfig:asymm_symm_icsi_fbl}]{%
		\includegraphics[width=0.95\figurewidth]{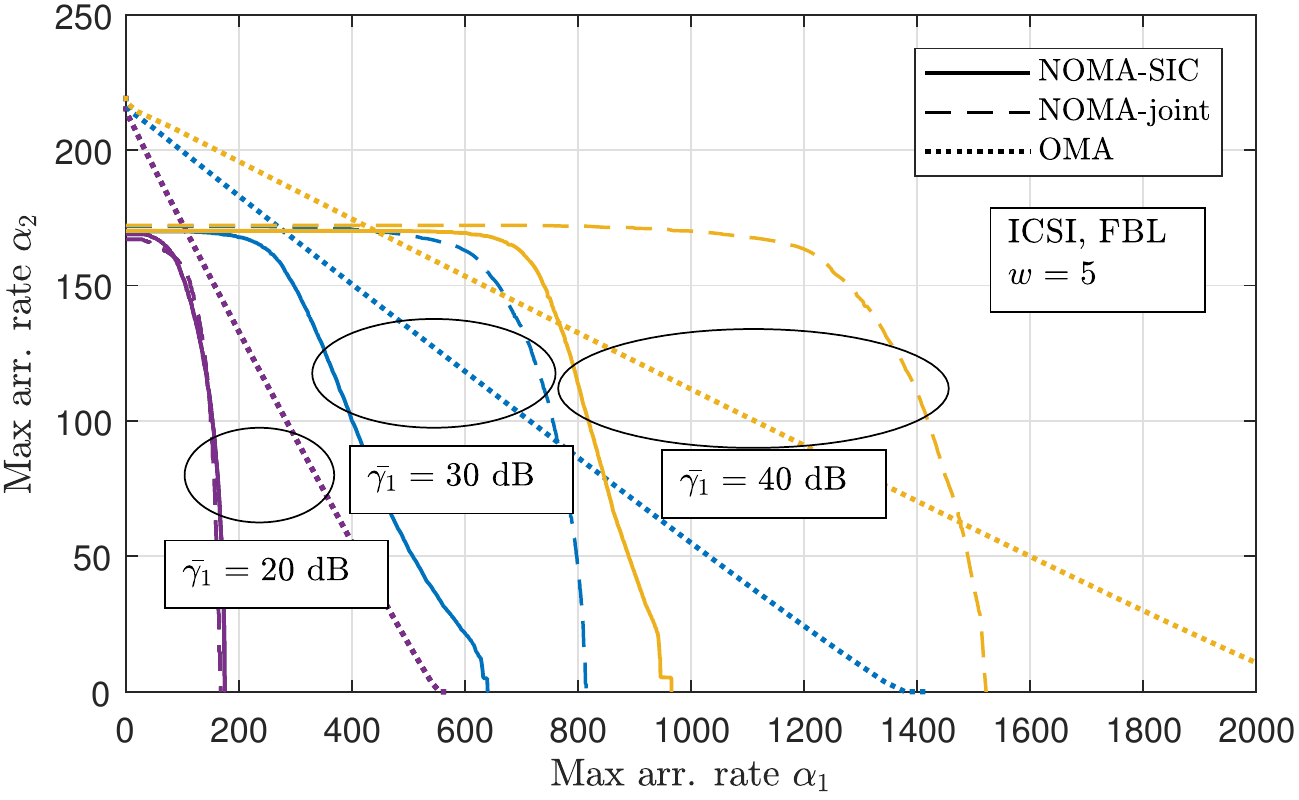}
	}
	\vfill
	\caption{Max arrivals $\alpha_1$, $\alpha_2$ in bits/slot s.t. $\pv(w)<10^{-8}$, $w=5$. Average SNR: $\snravgoma_1\in\{20,30,40\}~\mathrm{dB}$, $\snravgoma_2=15~\mathrm{dB}$. $\ntotal=250$, $\ntk{1}=\ntk{2}=25$, $\nd=200$. Imperfect CSI. a) Shannon capacity model (infinite blocklength) b) Finite blocklength model}
	\label{fig:asymm_symm7}
\end{figure}

\subsection{Different Channel Symmetries and Finite-Length Coding}
\label{ssec:numerics_fbl}
While we have previously considered a setup with $\snravgoma_1=30~\mathrm{dB}$, $\snravgoma_2=15~\mathrm{dB}$, we investigate in Fig.~\ref{fig:asymm_symm7} the cases $\snravgoma_1\in\{20,30,40\}~\mathrm{dB}$ and $\snravgoma_2=15~\mathrm{dB}$, i.e., we investigate different ratios between the users' average SNR values. Furthermore, we will now consider also finite blocklength effects. In Fig.~\ref{subfig:asymm_symm_icsi}, we show again results for the semi-realistic model with imperfect CSI, but still assuming that errors occur only when the rates are above the Shannon capacity (i.e., infinite blocklength), whereas Fig.~\ref{subfig:asymm_symm_icsi_fbl} shows results for the realistic model where finite blocklength effects are also considered. 
For $\snravgoma_1=30~\mathrm{dB}$, the results in Fig.~\ref{subfig:asymm_symm_icsi} were already shown in Fig.~\ref{subfig:asymm_icsi}. When comparing them to the new results in Fig.~\ref{subfig:asymm_symm_icsi_fbl}, we find that NOMA-joint still outperforms OMA, but NOMA-SIC performs worse than OMA when finite-length coding is taken into account.
For $\snravgoma_1=40~\mathrm{dB}$, NOMA-joint outperforms both NOMA-SIC and OMA by a large margin. \hlchange{In this scenario where the difference between the two users' average SNR is large, NOMA-SIC can still outperform OMA, but only by a fairly small margin.}
For $\snravgoma_1=20~\mathrm{dB}$, the chosen power allocation $\powfac_1=0.2$, $\powfac_2=0.8$ results in an almost symmetric scenario in terms of average SNR values \hlchange{(NOMA usually performs best in asymmetric scenarios, but we confirmed for this case that NOMA still exceeds OMA in the sum ergodic capacity).} 
We observe that OMA outperforms both NOMA schemes, and that there is a significant margin between OMA and the NOMA schemes once finite blocklength effects are considered. This indicates that the NOMA schemes suffer more from finite blocklength coding than the OMA scheme. 
It must be noted that the results for finite-length coding from \cite{scarlett2017dispersion} and \cite{molavianjazi2014phdthesis} are approximations, and we are not aware of information-theoretic bounds that can be used to verify their accuracy. Furthermore, the codewords for OMA are below 200 symbols long, so that \eqref{eq:rate_polyanskiy} starts to become inaccurate. However, it is noteworthy that finite blocklength effects seem to create a larger penalty for the NOMA schemes than for OMA, despite the fact that OMA operates with two codewords of shorter blocklength. 
This demonstrates that the blocklength itself is not the only factor that determines the performance impact of finite blocklength channel coding. 

%% file: Conclusions.tex

\section{Conclusions}
\label{sec:conclusions}
In this work, we analyzed the delay performance of NOMA in the uplink. We found that even under realistic assumptions, NOMA may be suitable for ultra-reliable low-latency communications, but only when joint decoding is used and only when there is a large difference between the two users' average SNR values.
However, joint decoding may be difficult to implement in practice. With SIC decoding, NOMA often performs worse than OMA when considering low-latency communications with more realistic system effects.

Aside from the interference-cancellation techniques investigated in this paper, simultaneous uplink from several users can also be enabled through multi-antenna technology, where the interference from different users can be mitigated through receive beamforming. Combining these two approaches would yield an interesting extension to our results.